\documentclass{aa}

\usepackage{graphicx}
\usepackage{txfonts}
\usepackage{hyperref}
\usepackage{placeins}

\makeatletter
\renewcommand*{\@fnsymbol}[1]{\ensuremath{\ifcase#1\or *\or \dagger\or \ddagger\or
    \mathsection\or \mathparagraph\or \|\or **\or \dagger\dagger
    \or \ddagger\ddagger \else\@ctrerr\fi}}
\makeatother

\begin{document}

   \title{EP-FXT observations of the cool-core cluster Abell 478 out to $R_{200}$: Thermodynamic properties and azimuthal asymmetry}


   \author{J.X. Sun \inst{1,2}
          \and
          Y. Chen \inst{2}\thanks{Corresponding author:\email{ychen@ihep.ac.cn}}
          \and
          S.M. Jia \inst{2}
          \and
          C.K. Li \inst{2}
          \and
          J. Zhang \inst{2}
          \and
          X.J. Yang \inst{1}\thanks{Corresponding author:\email{xjyang@xtu.edu.cn}}
          \and
          H. Yu \inst{3}
          \and
          A. Liu \inst{3}
          \and
          X.Y. Zheng \inst{3}
          \and
          W.W. Cui \inst{2}
          \and
          D.W. Han \inst{2}
          \and
          H.S. Zhao \inst{2}
          \and
          X.F. Zhao \inst{2}
          \and
          J.J. Xu \inst{2}
          }

   \institute{Hunan Key Laboratory for Stellar and Interstellar Physics and School of Physics and Optoelectronics, Xiangtan University, Xiangtan, Hunan, 411105, China
         \and
             State Key Laboratory of Particle Astrophysics, Institute of High Energy Physics, Chinese Academy of Sciences, Beijing 100049, China
         \and
             School of Physics and Astronomy, Beijing Normal University, Beijing 100875, China}
         
   \date{Received 21 April, 2026; accepted 9 September, 2026}

   \titlerunning{EP-FXT observations of Abell 478 out to $R_{200}$}
   \authorrunning{J.X.Sun et al.}

  \abstract
{The outskirts of galaxy clusters are key regions for studying their thermodynamic evolution and for testing the validity of the hydrostatic equilibrium assumption. However, because the gas surface brightness in the outskirts is extremely faint, precise measurements of these properties have long been limited by the instrumental background, field-of-view coverage, and observational depth. Furthermore, even in globally regular cool-core clusters, local azimuthal thermodynamic asymmetries can affect mass estimates. }
{We use deep observations from the \textit{Einstein} Probe Follow-up X-ray Telescope (EP-FXT) to systematically investigate the gas distribution and thermodynamic properties of the cool-core galaxy cluster Abell~478 (A478) from the central region out to $R_{200}$ and to explore its azimuthal asymmetry and the impact of local dynamical disturbances on mass estimations.}
{Benefiting from the low background, large field of view, and imaging spectroscopic capability of EP-FXT, we derived the surface-brightness and radial temperature distributions of A478, and further inferred the electron density, thermal pressure, entropy, and total mass profiles. The surface-brightness distribution was fitted with a double-$\beta$ model. The total mass profile was calculated under the assumption of hydrostatic equilibrium and parameterized with a Navarro--Frenk--White (NFW) model. In addition, we performed independent thermodynamic analyses of the NE and SW sectors to investigate local azimuthal structural differences.}
{The surface-brightness distribution of A478 is well described by a double-$\beta$ model, and its temperature profile exhibits the characteristic behavior of a cool-core cluster, with a cool center, a rise in temperature toward larger radii, and a gradual decline in the outskirts. The total mass profile derived under the hydrostatic equilibrium assumption is well fitted by the NFW model, which yields $R_{200}=2082\pm95\ \mathrm{kpc}$ and $M_{200}=(1.12\pm0.15)\times10^{15}\ \mathrm{M}_{\odot}$; this indicates that the system is close to quasi-static equilibrium on global scales. On the other hand, the sector analysis reveals clear azimuthal asymmetry: the SW direction exhibits lower temperatures, higher densities, and lower entropies in the intermediate and outer regions, together with observational signatures consistent with a cold-front candidate. Combined with the two-dimensional temperature distribution, these results suggest that the intracluster medium of A478 is likely affected by local dynamical processes such as gas sloshing.}
{A478 exhibits a dynamical state that is globally relaxed but locally disturbed. This work extends the thermodynamic measurements of the cluster out to $R_{200}$ and shows that even in a globally regular cool-core cluster, local nonequilibrium structures can still have a non-negligible impact on thermodynamic analyses and mass estimates.}


\maketitle

\section{Introduction}
   Galaxy clusters are the largest self-gravitating bound systems known in the Universe. They form through the growth of primordial density perturbations under gravity followed by nonlinear collapse, and therefore serve as important physical probes for studying cosmological parameters as well as the formation and evolution of large-scale structure \citep{2019SSRv..215....7W,2012ARA&A..50..353K}. Although the central regions of many galaxy clusters have approached a quasi-static equilibrium state, their outskirts continue to experience matter accretion and mergers along the surrounding filamentary structures. As a result, the outskirts constitute a transition zone between the quasi-static intracluster medium (ICM) in the inner regions and the collapsing material outside, making them key sites for investigating matter assembly, energy transport, and gas dynamical processes \citep{2017MNRAS.469.1476S}.

   However, the gas in cluster outskirts is extremely tenuous, and its X-ray surface brightness is usually 2--3 orders of magnitude lower than that in the central region. This places stringent requirements on the sensitivity, background control, and field-of-view coverage of X-ray telescopes. With the development of observational techniques based on X-ray emission, the Sunyaev--Zeldovich effect \citep{1980MNRAS.190..413S}, and optical weak lensing \citep{2001PhR...340..291B}, the study of cluster outskirts has become an important frontier in both galaxy-cluster astrophysics and precision cosmology \citep{2019SSRv..215....7W}. \textit{Suzaku}, \textit{XMM-Newton}, and \textit{Chandra} have played important roles in this field, extending measurements of the temperature, surface brightness, entropy, and hydrostatic equilibrium (HSE) mass of several nearby massive clusters to radii close to $R_{200}$ (i.e., the virial radius), including systems such as Abell~1795, Abell~2142, Abell~2029, and the X-COP cluster sample \citep{2009PASJ...61.1117B,2011PASJ...63S1019A,2012MNRAS.422.3503W,2017AN....338..293E,2018A&A...614A...7G,2019A&A...621A..41G,2019A&A...621A..39E}. For some intermediate- to high-redshift clusters with smaller angular sizes, a single or a few \textit{XMM-Newton} pointings can also trace the X-ray emission to nearly $R_{200}$, as in the cases of Abell~2218 and CL0016+16 \citep{2005A&A...433..777P,2007A&A...476...63S}. These studies have revealed several important phenomena in cluster outskirts, including declining temperature profiles, entropy profiles that deviate from self-similar expectations, gas clumping, nonthermal pressure support, and azimuthal asymmetries. Nevertheless, such observations are still affected to varying degrees by systematic uncertainties associated with the instrumental background, limited field of view, and mosaicking of multiple pointings. In recent years, the all-sky X-ray survey performed by the extended ROentgen Survey with an Imaging Telescope Array \textit{(eROSITA)} has significantly improved the observational coverage of the outer regions of galaxy clusters \citep{2021A&A...647A...1P}. However, its survey mode still limits the detailed characterization of faint thermal structures. 
   
   In this context, the \textit{Einstein} Probe (EP) Follow-up X-ray Telescope (FXT) provides new observational capabilities for studying the low-surface-brightness gas in cluster outskirts. The EP is the first Chinese focusing X-ray imaging astronomy satellite; it operates in a low-Earth orbit at an altitude of about 600 km \citep{2022hxga.book...86Y,2025SCPMA..6839501Y}. Its primary scientific goals are to monitor the dynamic X-ray sky with high sensitivity over a wide field of view and to perform rapid follow-up observations of newly discovered transient sources. The FXT (Chinese name: FengXingTian) on board EP \citep{2020SPIE11444E..5BC,2025SPIE13531E..02C} adopts a short-focal-length Wolter-I focusing optical design with a focal length of 1.6 m, and is equipped with two co-aligned modules and pixelated pn-junction charge-coupled device (pnCCD) focal-plane detectors. It operates over the 0.3--10 keV energy band \citep{2023ExA....55..603C,2025RAA....25a5002Z}, with a pixel scale of approximately $9.6\arcsec$ and a field of view of about $1^\circ$. FXT achieves a good balance between effective area, angular resolution, and field-of-view coverage, making it suitable not only for rapid and detailed observations of transient sources, but also particularly well suited for imaging and spectroscopic studies of extended low-surface-brightness sources such as galaxy clusters. Benefiting from its low-Earth-orbit environment and low-background design, the in-orbit particle-induced background of FXT is significantly reduced \citep{2022APh...13702668Z}. Current in-orbit calibration results show that its instrumental background is about one order of magnitude lower than that of \textit{XMM-Newton} \citep{2006SPIE.6266E..42Y}, and only about one-fifth that of eROSITA \citep{2021A&A...647A...1P,2025RAA....25k5019Z}. This low-background performance gives EP-FXT an advantage in detecting faint diffuse emission in cluster outskirts and near the virial radius \citep{2025A&A...700A.248Z,2026A&A...713A.110T}. At the same time, the relatively large field of view of FXT enables a continuous region from the cluster center to large radii to be covered within a single pointing, thereby reducing the systematic uncertainties introduced by mosaic observations.

   The ICM is the dominant baryonic component of galaxy clusters, and its thermal emission is primarily emitted in the X-ray band \citep[e.g.,][]{2025SCPMA..6839501Y}. Therefore, precise measurements of the spatial distribution and thermodynamic state of the ICM are essential for understanding the formation history of galaxy clusters, feedback processes, and their gravitational equilibrium structure. In this work, we selected the typical cool-core galaxy cluster A478 \citep{1999ApJS..125...35S} as our target. A478 has long been regarded as a relatively regular cool-core system with prominent central cooling \citep{2005ApJ...630..191S,2008A&A...479..307B}. The thermal structure, cooling properties, and metal-abundance distribution of its core region have been extensively studied by EXOSAT \citep{1991MNRAS.252..414E}, \textit{Einstein} \citep{1995MNRAS.273...72W}, Ginga \citep{1992MNRAS.255..431J}, ROSAT \citep{1994MNRAS.269..589W}, \textit{Chandra} \citep{2003ApJ...587..619S}, and \textit{XMM-Newton} \citep{2004A&A...423...33P}. In addition, high-resolution X-ray and radio observations have revealed complex structures in its core associated with the central active galactic nucleus, including hot spots, X-ray cavities, and diffuse radio emission, indicating that the system is not completely quiescent and may be jointly influenced by active galactic nucleus feedback and local dynamical processes \citep{2003ApJ...587..619S,2005ApJ...630..191S}. However, previous studies have mainly focused on the central and intermediate-radius regions, and systematic constraints on the outskirts from $0.5R_{200}$ to around $R_{200}$ remain limited, particularly for the temperature and other thermodynamic properties. Since the outskirts are not only the transition zone connecting the cluster to the surrounding large-scale structure but also an important window for testing the assumption of HSE and investigating variations in the baryon distribution \citep{2010ApJ...721.1105B,2014PhyU...57..317V}, it is therefore necessary to carry out a larger-scale systematic analysis of A478 using a new generation of observations better suited to the study of diffuse, low-surface-brightness emission.

   On the other hand, a growing body of observations has shown that even in cool-core galaxy clusters with globally regular morphologies, the thermodynamic distribution of the ICM often exhibits significant azimuthal asymmetry, that is, systematic differences in temperature, density, and entropy along different directions. Such asymmetric structures may arise from anisotropic active galactic nucleus feedback \citep{2016ApJ...818..181Y, 2017MNRAS.472.4707B} and are also often associated with gas sloshing and cold-front activity triggered by minor mergers \citep{2007PhR...443....1M, 2010ApJ...717..908Z}. In some typical cool-core clusters, the observed spiral-like cool-gas structures and non-radial disturbances reveal the important role of core dynamics in shaping the thermodynamic structure of the ICM \citep{2005MNRAS.356.1022S,2010MNRAS.405.1624M}. Identifying such local thermodynamic deviations is important not only for understanding the dynamical evolution of cool-core clusters, but also for assessing whether mass estimates based on the assumption of HSE are affected by local nonequilibrium processes. For A478, previous two-dimensional temperature studies have already suggested significant inhomogeneities in different azimuthal directions \citep{2008A&A...479..307B}, indicating that although the system appears globally regular, it can still preserve thermodynamic imprints left by local disturbances.

   Using deep EP-FXT observations, we systematically investigated the gas distribution and thermodynamic properties of A478 from its cool-core center out to $R_{200}$. Benefiting from the low background, large field of view, and imaging spectroscopic capability of EP-FXT, we derived the temperature, electron density, thermal pressure, entropy, and total mass profiles out to $R_{200}$. We further performed sector analyses in the NE and SW directions to explore the local thermodynamic asymmetry and its possible dynamical origin. 

Section~\ref{sec:obs2} describes the observations and data reduction procedures. Section~\ref{sec:obs3} presents the results of the global radial analysis of A478, including the surface-brightness distribution, the projected and deprojected temperature profiles, the electron-density distribution, and the pressure, entropy, and hydrostatic mass profiles derived from the deprojected temperature and electron density. Section~\ref{sec:obs4} gives the azimuthally resolved thermodynamic results for the NE and SW sectors. Section~\ref{sec:obs5} discusses the physical implications of the local thermodynamic asymmetry and compares our results with previous studies. Section~\ref{sec:obs6} summarizes the main conclusions of this work.

   Unless otherwise stated, we adopt a flat $\Lambda$ cold dark matter (CDM) cosmology with $H_0 = 70\ \mathrm{km\ s^{-1}\ Mpc^{-1}}$, $\Omega_{\mathrm{m}} = 0.3$, and $\Omega_{\Lambda} = 0.7$. Under this cosmology, at the redshift of A478, $z = 0.0881$, $1\,\mathrm{arcmin}$ corresponds to $99\,\mathrm{kpc}$. Throughout this paper, the term ``virial radius'' refers to $R_{200}$, defined as the radius within which the mean enclosed density is 200 times the critical density of the Universe at the cluster redshift.

\section{Observations and data reduction}
   \label{sec:obs2}
      
   \subsection{Observations}
   \label{sec:observations}

   This study is based on multiple on-axis EP-FXT observations of the galaxy cluster A478. The observations were carried out between 2024 October 26 and 2024 December 6, comprising a total of 14 valid observations with a combined exposure time of approximately 200 ks. The basic information for each observation is listed in Table~\ref{tab:obslog}. All observations were obtained in full frame mode \citep{2024PASP..136j5001Z}, in which the entire imaging area is read out periodically. This mode is well suited to extended sources such as galaxy clusters, which have large angular extents and show no significant variability on observational timescales. Before scientific analysis, the raw telemetry data were first processed through the standard preprocessing procedure, including the identification and removal of bad and flickering pixels, event grade assignment, pulse-invariant calculation, and subsequent event screening \citep{2025RDTM....9..237J}.

   The data reduction was carried out mainly using the FXT Data Analysis Software (\texttt{FXTDAS}, version 1.30) \footnote{\href{http://epfxt.ihep.ac.cn/analysis/}{http://epfxt.ihep.ac.cn/analysis/}}\citep{zhao2025data}. For each observation, we applied the standard pipeline tool \texttt{fxtchain} to process the raw event data in a uniform manner. This procedure automatically performs the key steps required for scientific analysis, including energy calibration, event reconstruction, particle-event identification and screening, and the generation of good time intervals, and produces the standardized event files used in the subsequent analysis. On this basis, we further extracted imaging products, spectral products, and the corresponding auxiliary files for subsequent image processing, surface-brightness analysis, and spatially resolved spectral analysis. In addition to \texttt{FXTDAS}, we also made use of the general high-energy astrophysics software packages \texttt{HEASoft} (version 6.33.1) \footnote{\href{https://heasarc.gsfc.nasa.gov/lheasoft/}{https://heasarc.gsfc.nasa.gov/lheasoft/}} and \texttt{CIAO} (version 4.16)\footnote{\href{https://cxc.cfa.harvard.edu/ciao/}{https://cxc.cfa.harvard.edu/ciao/}} for parts of the image processing, region manipulation, and cross-checks, in order to ensure the consistency of the overall analysis procedure and the robustness of the results.

  \subsection{Image processing and background modeling}
  Based on the event files processed through the standard pipeline for each observation, we first extracted count images in the 0.3--7.0 keV band and generated the corresponding exposure maps. The \texttt{fxtchain} pipeline can output exposure maps both with and without vignetting correction; the former were used for the subsequent exposure correction in the imaging analysis. To avoid the influence of low-exposure regions near the detector edges and local discontinuous structures on the merged image, we retained only those pixels with exposure times higher than 20\% of the maximum exposure of each observation, and constructed a uniform mask accordingly. We then used \texttt{fxtootest} to identify out-of-time events, and combined \texttt{ftcopy} and \texttt{ftimgcalc} to extract and correct the 0.3--7.0 keV images, subtracting the out-of-time component from the count image of each individual observation. After applying the mask, we obtained the net count image and the high-quality exposure map for each observation. Considering the slight differences in pointing and attitude among the individual observations, we further used \texttt{reproject\_image} in \texttt{CIAO} to reproject the count images and exposure maps onto a common sky coordinate grid, and then co-added them separately to produce the total count image and total exposure map.

\begin{figure}[htbp]
    \centering
    \includegraphics[width=0.48\textwidth]{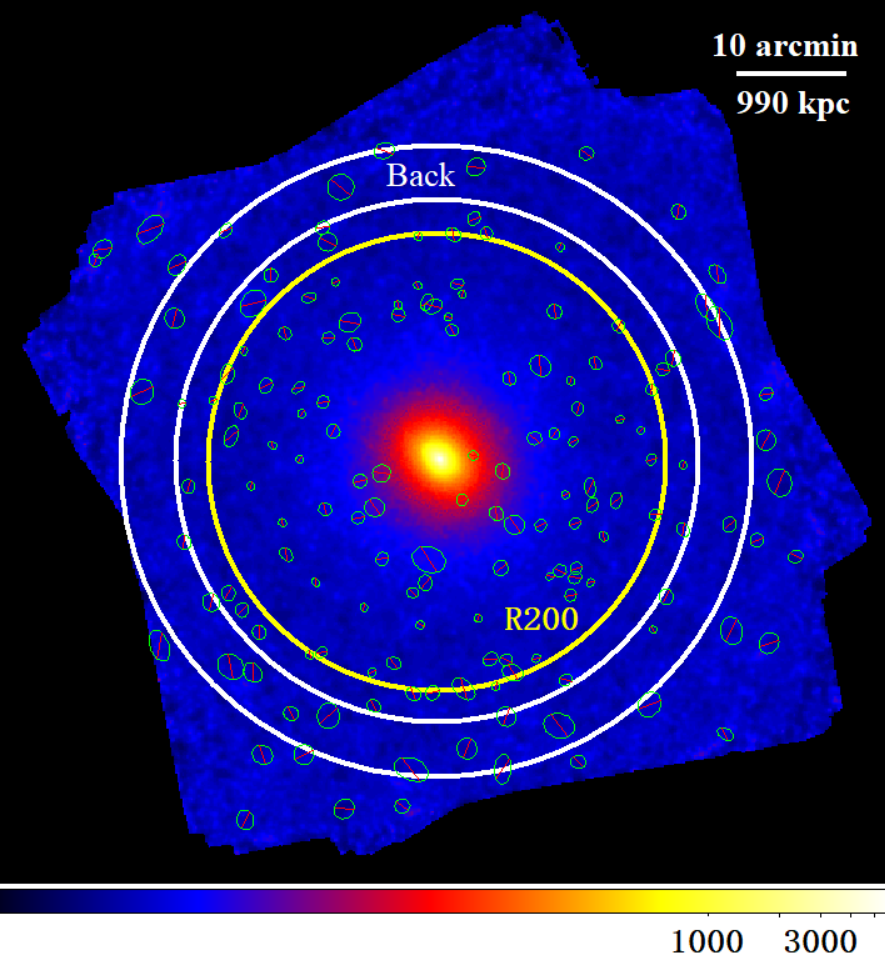}
    \caption{Merged EP-FXT X-ray image of A478 in the 0.3--7.0 keV band. The image has been processed with point-source detection and removal, exposure correction, and adaptive smoothing. The yellow circle marks the $R_{200}$ radius adopted in this work. The region between the two white annuli indicates the local background extraction region. The green ellipses denote the identified point-source masks. The scale bar corresponds to 10 arcmin, or approximately 990 kpc.
    }
    \label{fig:epfxt_a478_images}
\end{figure}

  To suppress the contamination from discrete sources in the imaging analysis of the diffuse emission, we used \texttt{wavedetect} in \texttt{CIAO} on the merged image to identify point sources within the field of view, and excluded them from the subsequent analysis. The masked point-source regions were then filled using a pixel reconstruction method based on Poisson statistics, in order to preserve the spatial continuity and local statistical properties of the diffuse emission as much as possible. We subsequently corrected the total count image using the vignetting-corrected total exposure map, and normalized it by the maximum exposure time to obtain the exposure-corrected image in the 0.3--7.0 keV band. Finally, we applied \texttt{dmimgadapt} to perform adaptive smoothing on the image, producing the final X-ray image of A478 used in this work (Fig.~\ref{fig:epfxt_a478_images}). Overall, the X-ray emission of A478 appears relatively regular in morphology, with a pronounced enhancement in surface brightness toward the center. These morphological features will be further discussed in Sect.~\ref{sec:sb} in combination with the surface-brightness distribution.

  For the subsequent spatially resolved spectral analysis, we constructed the background separately by considering the two main components: the non-X-ray background (NXB) and the cosmic X-ray background (CXB). Owing to the low particle background level of EP-FXT, the NXB in the imaging area can be estimated using events recorded in the pnCCD storage area. To obtain the local sky background, we selected an annular region at radii of 25--29 arcmin in the outer field of view as the background extraction region (shown by the white annuli in Fig.~\ref{fig:epfxt_a478_images}), while avoiding obvious source contamination as much as possible in order to minimize the residual contribution of cluster emission in the background region. Since the current standard \texttt{FXTDAS} products do not yet provide a complete background-modeling module suited to the requirements of this work, we constructed the background spectrum separately for each source region. Specifically, the storage-area events were first processed with \texttt{fxtchain} in \texttt{datatype=fsaevt} mode, and the particle background spectrum corresponding to the imaging area was then generated using \texttt{fxtbkggen} \citep{2025RAA....25k5019Z}. The resulting spectrum was scaled by the effective area of the extraction region to obtain the NXB component for each source region and for the background region. We then extracted the total background spectrum from the background annulus, and obtained the local CXB component by subtracting the corresponding NXB contribution. This CXB spectrum was subsequently corrected for vignetting using the exposure map, and converted into the CXB spectrum for each source region by taking into account the differences in area and effective exposure between the source and background regions. Finally, the background spectrum for each source region was obtained by adding its corresponding NXB component and the corrected CXB component. Region definition and spectral extraction were carried out using \texttt{DS9} (version 5.2) \footnote{\href{https://sites.google.com/cfa.harvard.edu/saoimageds9}{https://sites.google.com/cfa.harvard.edu/saoimageds9}} and \texttt{Xselect} (version 2.5b) \footnote{\href{https://heasarc.gsfc.nasa.gov/docs/software/ftools/xselect/xselect.html}{https://heasarc.gsfc.nasa.gov/docs/software/ftools/xselect/xselect.html}}, respectively.

  For the error treatment, we propagated the statistical uncertainties through each step of the background construction procedure, including the counting-statistics error, the uncertainty associated with NXB subtraction, the uncertainty introduced by area-scaling, and the uncertainty arising from the vignetting correction. The resulting total background error spectrum was written in a format compatible with \texttt{XSPEC} (version 12.14.1) \footnote{\href{https://heasarc.gsfc.nasa.gov/docs/xanadu/xspec/manual/manual.html}{https://heasarc.gsfc.nasa.gov/docs/xanadu/xspec/manual/manual.html}} and was included in the subsequent spectral fitting, thereby ensuring the reliability and statistical robustness of the spectral analysis in the low-surface-brightness outer regions.

\section{Global radial thermodynamic properties}
   \label{sec:obs3}
   After point-source removal, exposure correction, and background construction for the EP-FXT images, we further investigated the gas distribution and thermodynamic properties of A478 on radial scales. In this section we first extract and fit the radial surface-brightness profile from the 0.3--2.0 keV image, in order to constrain the spatial distribution of the gas emission. We then derive the projected temperature profile through annular spectral analysis and further perform a geometric deprojection to obtain the shell temperatures used for the calculation of three-dimensional thermodynamic quantities. On this basis, we combine the electron-density profile inferred from the deprojected surface-brightness distribution with the deprojected temperature profile to derive the pressure, entropy, and hydrostatic mass distributions. The radial results are presented below.   

   \subsection{Surface brightness profile}
   \label{sec:sb}
   To investigate the radial structural properties of A478, we extracted its X-ray surface-brightness distribution from the merged 0.3--2.0 keV image after point-source removal and exposure correction. This energy band was chosen mainly because it provides both a relatively high signal-to-noise ratio and a relatively low background level, allowing a more robust tracing of the spatial distribution of the diffuse thermal gas in the cluster. The surface-brightness profile was extracted using \texttt{Sherpa} (version 4.16.0) \footnote{\href{https://cxc.cfa.harvard.edu/sherpa/}{https://cxc.cfa.harvard.edu/sherpa/}}, with the cluster center defined as the peak of the X-ray emission, and the radial range extending to about 30 arcmin. Specifically, we extracted the radial profile using the \texttt{get\_data\_prof} tool in Sherpa, with the radial step set to \texttt{rstep=2}, corresponding to a bin width of two image pixels for each radial bin. Given that the pixel scale of the EP-FXT image is approximately $9.6\arcsec$, this setup corresponds to a radial sampling interval of about $19.2\arcsec$, or $0.32\arcmin$. It should be noted that this value represents the sampling scale of the surface-brightness profile, whereas the actual angular resolution in the central region is determined primarily by the point spread function (PSF) of EP-FXT.

   The radial surface-brightness profile extracted with Sherpa corresponds to the area-normalized mean surface brightness within each annulus, that is, an area-weighted mean surface-brightness profile. To test whether the mean surface brightness could overestimate the gas density in the outskirts because of gas clumping or residual small-scale structures, we further performed a mean--median comparison. Inspired by the method of \citet{2015Natur.528..105E}, we divided each radial annulus into 12 azimuthal sectors and, using the same point-source and exposure masks as in the main analysis, constructed both the mean and azimuthal-median surface-brightness profiles after subtraction of the CXB+NXB components. We then fitted the two profiles with the same double-$\beta$ model. Although the individual parameters of the double-$\beta$ model show some degeneracies, the total surface-brightness models obtained with the two methods are very similar. Over the full fitted radial range, the median model ratio is $S_{\mathrm{mean,model}}/S_{\mathrm{median,model}}=1.027$. Since, for a given temperature and metal abundance, $n_{\mathrm e}\propto S_{\mathrm X}^{1/2}$, this corresponds to an electron-density ratio of $n_{\mathrm{e,mean}}/n_{\mathrm{e,median}}=1.014$. This indicates that using the mean surface brightness rather than the azimuthal-median surface brightness affects the recovered gas-density profile by only 1.4\%. We therefore retain the area-weighted mean surface-brightness profile as the reference profile in the main analysis, and treat the mean--median difference as a small systematic uncertainty in the normalization of the outer density profile.   
 
   To quantitatively describe the composite structure of the central cool core and the outer main halo in A478, we fitted the surface-brightness profile with the classical double-$\beta$ model, written as

\begin{equation}
S(r) = \sum_{i=1}^{2} S_i \left[1 + \left( \frac{r}{r_{c,i}} \right)^2 \right]^{-3\beta_i + 0.5} + b_0
\label{eq:double_beta}
,\end{equation}
where $S_i$ is the surface-brightness normalization of the $i$th component, $r_{\mathrm{c},i}$ is the core radius, $\beta_i$ is the outer slope parameter, and $b_0$ represents a constant background term. For the observed surface-brightness profile without explicit background subtraction, $b_0$ represents the total background level of the CXB+NXB. For the surface-brightness profile after subtraction of the NXB and CXB components, $b_0$ is used only to describe any possible residual constant background component.

\begin{figure}[htbp]
    \centering
    \includegraphics[width=0.48\textwidth]{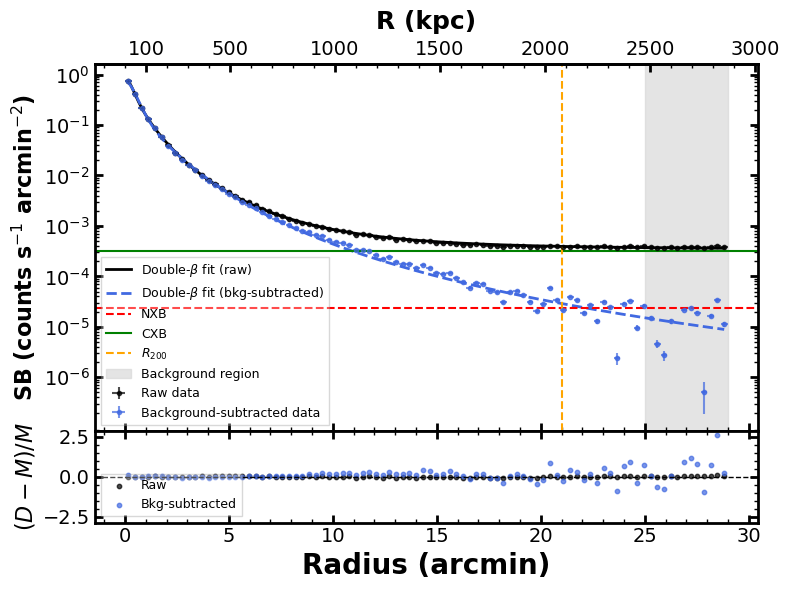}
    \caption{Radial surface-brightness profile of A478 in the 0.3--2.0 keV band. The black points represent the observed profile without explicit background subtraction, and the solid black line shows the corresponding best-fitting double-$\beta$ model. The blue points represent the surface-brightness profile after subtraction of the NXB and CXB components, and the dashed blue line shows the corresponding best-fitting model. The dashed red line (solid green) denotes the NXB (CXB) component. The vertical dashed orange line marks the position of $R_{200}$, and the gray shaded region indicates the outer annulus used for background estimation. The lower panel shows the relative residuals, $(D-M)/M$, where $D$ and $M$ denote the observed data and the model prediction, respectively.}
    \label{fig:sb_profile}
\end{figure}

\begin{table}[htbp]
\centering
\caption{Best-fitting double-$\beta$ model parameters for the radial surface-brightness profile of A478.}
\label{tab:beta_fit}
\begin{tabular}{lc}
\hline\hline
Parameter & Value \\
\hline
        $\beta_1$ & $0.74 \pm 0.03$ \\
        $\beta_2$ & $0.81 \pm 0.02$ \\
        $r_{\mathrm{c1}}$ (arcmin) & $0.74 \pm 0.02$ \\
        $r_{\mathrm{c2}}$ (arcmin) & $2.80 \pm 0.12$ \\
        $s_1$ ($10^{-1}\ \mathrm{counts\ s^{-1}\ arcmin^{-2}}$) & $6.86 \pm 0.10$ \\
        $s_2$ ($10^{-2}\ \mathrm{counts\ s^{-1}\ arcmin^{-2}}$) & $5.62 \pm 0.01$ \\
        $b_0$ ($10^{-5}\ \mathrm{counts\ s^{-1}\ arcmin^{-2}}$) & $1.17 \pm 0.12$ \\
\hline
\end{tabular}

\tablefoot{The fit is performed in the 0.3--2.0 keV band after subtraction of the NXB and CXB components. The uncertainties correspond to $1\sigma$.}
\end{table}
\vspace{\baselineskip}
   Figure~\ref{fig:sb_profile} presents the double-$\beta$ model fit to the radial surface-brightness profile of A478, and Table~\ref{tab:beta_fit} lists the corresponding best-fitting parameters together with their $1\sigma$ uncertainties. The surface brightness of A478 shows a pronounced enhancement in the central region and decreases rapidly with increasing radius, exhibiting the characteristic radial distribution of a typical cool-core galaxy cluster. The double-$\beta$ model provides a good description of this surface-brightness structure, which consists of both a compact central cool-core component and an extended large-scale main-halo component. In particular, the core radius of the outer component, $r_{\mathrm{c},2} = 2.800 \pm 0.120$ arcmin, is significantly larger than that of the inner component, $r_{\mathrm{c},1} = 0.737 \pm 0.024$ arcmin, indicating that the second component dominates the surface-brightness distribution on larger spatial scales. Meanwhile, $\beta_2 = 0.811 \pm 0.016$ is slightly larger than $\beta_1 = 0.743 \pm 0.033$, suggesting that the outer component itself has a somewhat steeper radial decline. Overall, the surface-brightness profile of A478 shows both a clear central concentration and a regular, smooth extended distribution on larger scales, consistent with its global morphology as a relatively relaxed cool-core galaxy cluster. Furthermore, an estimate based on the expected surface-brightness level of the best-fitting double-$\beta$ model in the 25--29 arcmin background annulus shows that the residual projected cluster emission in this region accounts for only about 3\% of the total signal. This indicates that the background annulus adopted in this work is sufficiently clean for the subsequent analysis.

   \subsection{Annular spectral analysis}
   \label{sec:temperature}
   To investigate the radial thermal structure of A478, we performed an annular spectral analysis based on the EP-FXT observations. Centered on the peak of the X-ray emission, we defined a central circular region with a radius of 0.4 arcmin, and further constructed ten concentric annuli outside it, with the outermost boundary extending to 21 arcmin, thereby covering the cluster out to $R_{200}$. This binning scheme was designed by jointly considering the spatial resolution, photon statistics, and the requirements of the subsequent derivation of thermodynamic quantities, thus providing a basis for constructing a robust radial temperature profile. For each observation, we extracted the source and background spectra separately from each annular region and generated the corresponding response files. The ancillary response files were produced using \texttt{fxtarfgen} with \texttt{extend=1} in the extended-source mode, while the redistribution matrix files were generated using \texttt{fxtrmfgen}. We then used \texttt{addspec} to combine the multiple observations within the same radial bin in a weighted manner, thereby obtaining, for each annulus, the final merged source spectrum used for fitting together with its corresponding background spectrum.

   The spectral fitting was performed with \texttt{XSPEC} using a single-temperature thermal plasma model modified by neutral hydrogen absorption, \texttt{TBabs} $\times$ \texttt{APEC}, over the 0.5--7.0 keV energy band. Since the lower-energy channels are more strongly affected by the limited detector response and the relatively large uncertainty in background modeling, and are also more susceptible to contamination from the soft X-ray background, channels below 0.5 keV were excluded from the fitting. All spectra were regrouped with \texttt{grppha} prior to fitting, and the $\chi^2$ statistic was adopted for parameter estimation. The cluster redshift was fixed at $z = 0.0881$. The line of sight toward A478 is subject to relatively high Galactic absorption and exhibits clear spatial variation. Based on \textit{XMM-Newton} observations, \citet{2004A&A...423...33P} showed that the absorption column density varies significantly across different regions in this direction, and that the excess absorption is most likely associated with Galactic foreground material. Therefore, to reduce the impact of absorption uncertainties at low energies on the temperature measurements, and to alleviate parameter degeneracies in annuli with relatively low signal-to-noise ratios, we did not adopt a single Galactic $N_{\mathrm H}$ value for all annuli. Instead, we used radially varying fixed $N_{\mathrm H}$ values based on \citet{2004A&A...423...33P} during the spectral fitting. In addition, because the metal abundance in the outer annuli is only weakly constrained and is strongly coupled with both the temperature and the absorption column density, we fixed the abundance at $0.30\,Z_\odot$ from the seventh annulus outward ($r \approx 4.7$ arcmin) in order to improve the stability of the temperature measurements in the outskirts.

\begin{figure}[htbp]
    \centering
    \includegraphics[width=0.48\textwidth]{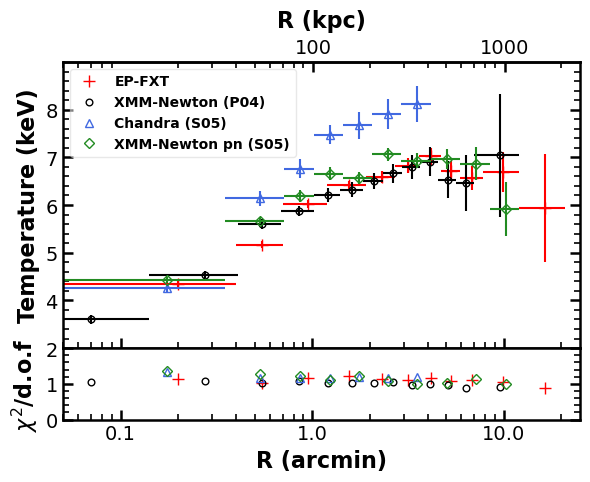}
    \caption{Radial projected temperature profile of A478. The red points show the temperature profile derived in this work from the EP-FXT data, the black circles the \textit{XMM-Newton} results reported by \citet{2004A&A...423...33P}, the blue triangles the \textit{Chandra} results reported by \citet{2005ApJ...630..191S}, and the green diamonds the \textit{XMM-Newton}/pn results from \citet{2005ApJ...630..191S}. The temperature error bars indicate the $1\sigma$ statistical uncertainties.}
    \label{fig:specfit}
\end{figure}

   The fitting results are shown in Fig.~\ref{fig:specfit}, and Table~\ref{tab:spectral_fit} lists the radial ranges and best-fitting parameters for each annulus. The radial temperature distribution of A478 exhibits the characteristic behavior of a cool-core galaxy cluster: the temperature is about 4.35 keV in the central region, then gradually increases with radius, and decreases to about 5.93 keV in the outer 12--21 arcmin annulus. Overall, this structure, characterized by a cool center, a temperature rise at intermediate radii, and a decline in the outskirts, is consistent with the typical radial temperature behavior of cool-core clusters. Figure~\ref{fig:specfit} also includes previously published projected temperature measurements from \textit{XMM-Newton} and \textit{Chandra} for comparison with our EP-FXT results. The \textit{XMM-Newton} temperature profile reported by \citet{2004A&A...423...33P} extends to about 12 arcmin, corresponding to approximately $0.5R_{200}$. The \textit{Chandra} temperature measurements of \citet{2005ApJ...630..191S} mainly cover the inner to intermediate radial range, while their \textit{XMM-Newton}/pn temperature profile extends further out to about 12 arcmin. The \textit{Chandra} results of \citet{2005ApJ...630..191S} are slightly higher than the EP-FXT and \textit{XMM-Newton} temperatures in part of the inner and intermediate regions, whereas the \textit{XMM-Newton} results of \citet{2004A&A...423...33P} and \citet{2005ApJ...630..191S} are broadly consistent with our EP-FXT results in terms of the overall radial trend. These differences may reflect systematic effects related to cross-calibration, energy-band selection, spatial resolution, and the treatment of absorption among different instruments. Furthermore, the EP-FXT data allow us to extend the temperature measurement into the outer 12--21 arcmin region, that is, close to the $R_{200}$ scale, thereby providing new observational constraints on the thermal structure of the outskirts of A478. This radial extension also provides the basis for the subsequent analysis of the thermodynamic properties and total mass distribution.

\begin{table}[htbp]
\centering
\footnotesize
\setlength{\tabcolsep}{4pt}
\caption{Results of the radial annular spectral fitting for A478.}
\label{tab:spectral_fit}
\begin{tabular}{ccc}
\hline\hline
Radial range & $kT$ & $\chi^2_{red}$ (d.o.f.) \\
(arcmin) & (keV) &  \\
\hline
$0.0$--$0.4$   & $4.35 \pm 0.07$ & 1.14 (996) \\
$0.4$--$0.7$   & $5.17 \pm 0.08$ & 1.03 (1058) \\
$0.7$--$1.2$   & $6.01 \pm 0.08$ & 1.16 (1154) \\
$1.2$--$1.9$   & $6.41 \pm 0.10$ & 1.22 (1166) \\
$1.9$--$2.7$   & $6.59 \pm 0.12$ & 1.12 (1113) \\
$2.7$--$3.6$   & $6.82 \pm 0.16$ & 1.18 (1041) \\
$3.6$--$4.7$   & $7.01 \pm 0.19$ & 1.16 (982) \\
$4.7$--$5.9$   & $6.70 \pm 0.21$ & 1.07 (866) \\
$5.9$--$7.8$   & $6.56 \pm 0.26$ & 1.11 (853) \\
$7.8$--$12.0$  & $6.69 \pm 0.42$ & 1.05 (968) \\
$12.0$--$21.0$ & $5.93 \pm 1.14$ & 0.89 (1268) \\
\hline
\end{tabular}

\tablefoot{The table gives the radial range, best-fitting temperature, reduced $\chi^2$, and number of degrees of freedom. The temperature uncertainties correspond to $1\sigma$.}
\end{table}

   \subsection{Deprojected thermodynamic distributions}

   Based on the surface-brightness and projected temperature analyses described above, we further reconstructed the three-dimensional thermodynamic structure of the gas in A478. The surface-brightness profile was used to constrain the electron-density distribution, while the projected temperature profile was converted into a three-dimensional shell temperature profile through geometric deprojection. By combining these two quantities, we further derived the pressure, entropy, and subsequent hydrostatic mass distributions, thereby providing a basis for characterizing the cool-core structure and the thermal state of the outskirts of A478.

\begin{figure}[htbp]
    \centering
    \includegraphics[width=0.48\textwidth]{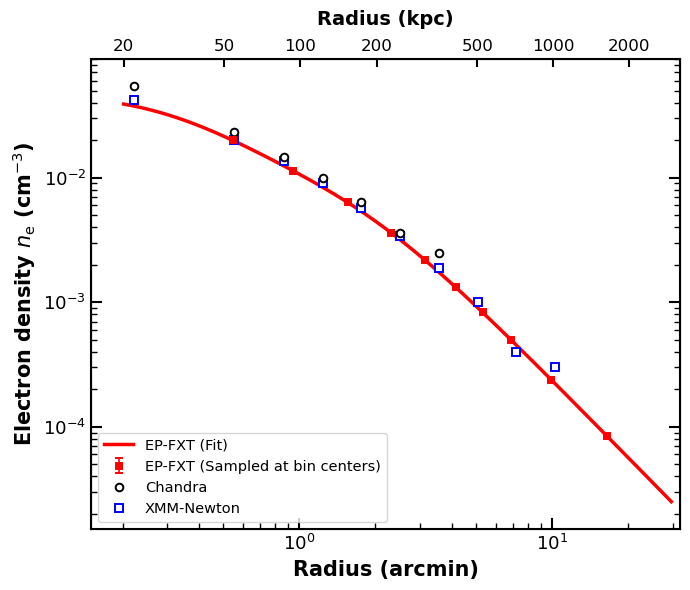}
    \caption{Radial electron-density profile of the galaxy cluster A478. The solid red line shows the best-fitting double-$\beta$ model derived from the deprojected EP-FXT surface-brightness profile. Since the uncertainty propagated from the statistical errors is smaller than the line width, the uncertainty band is not shown for clarity. The data points are plotted at the central radii of the corresponding temperature annuli, while the continuous curve represents the model distribution over the full radial range analyzed. The black circles and blue squares represent the electron-density measurements from the literature based on \textit{Chandra} and \textit{XMM-Newton}, respectively.}
    \label{fig:ne_fit}
\end{figure}

   To construct the three-dimensional electron-density distribution of A478, we reconstructed the radial profile of the volume emissivity from the background-subtracted 0.3--2.0 keV surface-brightness profile using a geometric deprojection method. In this procedure, we assumed that the cluster can be approximated globally as a spherically symmetric system. We then converted the volume emissivity into an electron-density distribution by adopting the same temperature and abundance settings as those used in the spectral analysis. Since the X-ray volume emissivity of a thermal plasma approximately satisfies $\epsilon \propto n_{\mathrm e}^2$, the radial electron-density profile, $n_{\mathrm e}(r)$, can be derived directly from the deprojected emissivity profile. It can be written as

\begin{equation}
n_{\mathrm{e}}(r) = n_{01}\left[1+\left(\frac{r}{r_{\mathrm{c1}}}\right)^2\right]^{-3\beta_1/2}
+ n_{02}\left[1+\left(\frac{r}{r_{\mathrm{c2}}}\right)^2\right]^{-3\beta_2/2}
,\end{equation}
where $n_{01}$ and $n_{02}$ represent the central electron densities of the inner and outer components, respectively; $r_{\mathrm{c}1}$ and $r_{\mathrm{c}2}$ are the corresponding core radii; and $\beta_1$ and $\beta_2$ are the slope parameters. We used the Python package \texttt{emcee} (version 3.1.6) \footnote{\href{https://emcee.readthedocs.io/en/v3.1.6/}{https://emcee.readthedocs.io/en/v3.1.6/}} to perform Markov chain Monte Carlo (MCMC) sampling in order to obtain the posterior distributions of the model parameters and their associated uncertainties. The best-fitting result is shown in Fig. \ref{fig:ne_fit}, and the corresponding parameters are listed in Table \ref{tab:ne_fit}. To illustrate the correlations among the double-$\beta$ model parameters and their uncertainties, we present the corner plot of the posterior distributions in Appendix~\ref{app:density_corner}. It should be noted that the uncertainties quoted for the electron-density profile mainly reflect the propagation of statistical errors associated with the surface-brightness fitting and geometric deprojection procedures, and do not explicitly include systematic uncertainties arising from background modeling, the assumption of spherical symmetry, or the adopted temperature and metal abundance. Therefore, these errors should be interpreted as statistical uncertainties. The fitting results show that the central normalizations of the inner and outer components are approximately $n_{01}=0.040\ \mathrm{cm^{-3}}$ and $n_{02}=0.008\ \mathrm{cm^{-3}}$, respectively. As an empirical parametrization, the double-$\beta$ model \citep{2003A&A...407...41C} provides a good description of the radial electron-density distribution of A478, simultaneously capturing the steep central density enhancement and the smoother radial decline at large radii, and thus provides a stable density model for the subsequent calculation of the pressure, entropy, and hydrostatic mass profiles.

   We further quantitatively assessed the potential impact of the EP-FXT PSF on the surface-brightness profile and the electron-density distribution. The on-axis PSF of EP-FXT has a half-power diameter of about $22\arcsec$--$24\arcsec$, which is substantially broader than the sub-arcsecond PSF of \textit{Chandra} and also slightly broader than the PSF of \textit{XMM-Newton} EPIC. For a strong cool-core cluster such as A478, which has a sharply peaked central surface-brightness distribution, PSF scattering can smooth the central surface-brightness gradient and may affect the central electron density and the related thermodynamic quantities. To evaluate this effect, we performed a forward test using the on-axis PSF model, in which the intrinsic double-$\beta$ surface-brightness model was convolved in two dimensions with the PSF. The corresponding intrinsic model was then geometrically deprojected using the same radial binning as adopted in the thermodynamic analysis. The results show that the PSF effect is mainly concentrated within the central $\sim1\arcmin$. After accounting for the PSF effect, the average electron density in the innermost radial bin increases by about 25\%, while the correction decreases rapidly with radius. Beyond $\sim1\arcmin$, the absolute change in the electron density is generally less than 5\%. Therefore, the PSF has a measurable effect on the density normalization in the very central cool-core region, but it does not significantly change the density distribution at intermediate radii and in the outskirts, nor does it affect the main conclusions of this work for the $R_{500}$--$R_{200}$ region. Since this test was performed only for the imaging surface-brightness distribution and does not include the energy-dependent PSF cross-talk between different spectral annuli, we retain the results without explicit PSF correction as the baseline measurements in this work, and use the above test as an estimate of the systematic uncertainty in the central region.
   
   To assess the reliability of the derived electron-density profile, we compared the EP-FXT results with the \textit{Chandra} and \textit{XMM-Newton} measurements reported by \citet{2005ApJ...630..191S}. Figure \ref{fig:ne_fit} shows the electron-density profile derived from the EP-FXT surface-brightness analysis together with the literature results. Overall, the EP-FXT, \textit{Chandra}, and \textit{XMM-Newton} electron-density profiles show consistent radial declining trends over the overlapping radial range. The \textit{Chandra} measurements are generally slightly higher than the EP-FXT results from the inner to intermediate radii, whereas the \textit{XMM-Newton} results are overall closer to the EP-FXT measurements. In light of the PSF test described above, the relatively broad EP-FXT PSF may be an important contributor to the density-normalization difference in the very central region, particularly for a strong cool-core cluster such as A478 with a sharply peaked central surface-brightness distribution, where PSF scattering can smooth the central surface-brightness gradient. Since the PSF effect decreases rapidly beyond about $1'$, the remaining differences at intermediate radii may also be related to differences in instrumental spatial resolution and analysis methods, rather than representing genuine physical differences in the gas distribution. The EP-FXT electron-density measurement further extends to large radii close to $R_{200}$, providing complementary constraints on the gas distribution in the outskirts of A478.

\begin{table}[htbp]
\centering
\caption{Best-fitting parameters of the double-$\beta$ model for the electron-density distribution of A478 and their $1\sigma$ uncertainties.}
\label{tab:ne_fit}
\begin{tabular}{lc}
\hline\hline
Parameter & Value \\
\hline
$\beta_1$ & $0.69 \pm 0.02$ \\
$\beta_2$ & $0.69 \pm 0.01$ \\
$r_{c1}$ (arcmin) & $0.38 \pm 0.01$ \\
$r_{c2}$ (arcmin) & $1.64 \pm 0.03$ \\
$n_{01}$ ($10^{-2}$ cm$^{-3}$) & $4.00 \pm 0.03$ \\
$n_{02}$ ($10^{-2}$ cm$^{-3}$) & $0.08 \pm 0.03$ \\
\hline
\end{tabular}
\end{table}

\begin{figure}[htbp]
    \centering
    \includegraphics[width=0.48\textwidth]{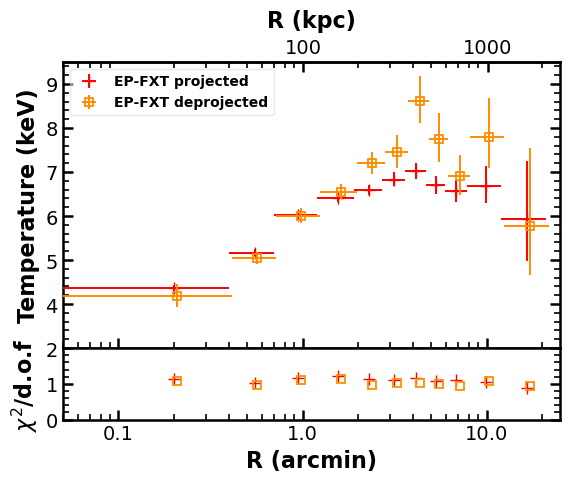}
    \caption{Radial projected and deprojected temperature profiles of A478 measured with EP-FXT. The projected temperatures derived from the annular spectral analysis are shown as red crosses, while the onion-peeling deprojected temperatures are shown as orange open squares. For clarity, the deprojected data points are shifted slightly to the right.}
    \label{fig:depro_fit}
\end{figure}

   After obtaining the electron-density profile, we further performed a geometric deprojection of the radial temperature distribution measured with EP-FXT, in order to derive the three-dimensional shell temperatures used in the subsequent calculation of the thermodynamic quantities. The deprojection was carried out using the same concentric annuli as adopted in the projected spectral analysis. Under the assumption of spherical symmetry, we constructed an onion-peeling geometric projection matrix to describe the contribution of each three-dimensional spherical shell to the spectra extracted from the different projected annuli. The projected contribution from the outer shells was then subtracted successively from the inner projected annuli, proceeding from the outside inward, to obtain the deprojected net spectrum corresponding to each shell. The deprojected spectra were then grouped according to their signal-to-noise ratios. The deprojected spectra were fitted using the same energy band and spectral model as those adopted in the projected spectral analysis. The redshift was fixed at $z=0.0881$, the hydrogen column density was fixed to the same radially varying values, and the treatment of the metal abundance was kept consistent with that used in the projected spectral fitting. Figure \ref{fig:depro_fit} shows the resulting deprojected temperature profile, together with the projected EP-FXT temperature profile for reference. Compared with the projected temperature profile, the deprojected temperature is lower in the central region and slightly higher at intermediate radii, as expected after partially removing the line-of-sight mixing of gas with different temperatures in a cool-core cluster. The deprojected temperature in the outermost shell is generally consistent with the projected temperature, and the small difference between the two mainly arises from the different statistical treatments of the two types of spectra. In this work, we combined this deprojected temperature profile with the electron-density model to calculate the pressure, entropy, and subsequent hydrostatic mass distributions. The statistical uncertainties were propagated to the derived physical quantities through Monte Carlo simulations.

   Based on the electron-density model and the deprojected temperature profile described above, we further calculated the radial distributions of the thermal pressure and entropy of A478. The thermal pressure reflects the ability of the ICM to provide thermal support against the gravitational potential of the cluster, and is therefore a key physical quantity in hydrostatic mass estimates \citep{2013SSRv..177..195R}. Entropy, on the other hand, records the thermal history and energy-injection processes of the gas, and is sensitive to the effects of radiative cooling, gravitational heating, and non-gravitational physical processes \citep{2005RvMP...77..207V}. In this work, we adopt the following definitions:
\begin{equation}
P_{\mathrm{e}} =kTn_{\mathrm{e}}
\label{eq:pressure}
\end{equation}
\begin{equation}
K_{\mathrm{e}} = kTn_{\mathrm{e}}^{-2/3}
\label{eq:entropy}
,\end{equation}
where $n_{\mathrm e}$ is the electron number density, $T$ is the gas temperature, and $k$ is the Boltzmann constant. 

\begin{figure*}[htbp]
\sidecaption
\includegraphics[width=12.8cm]{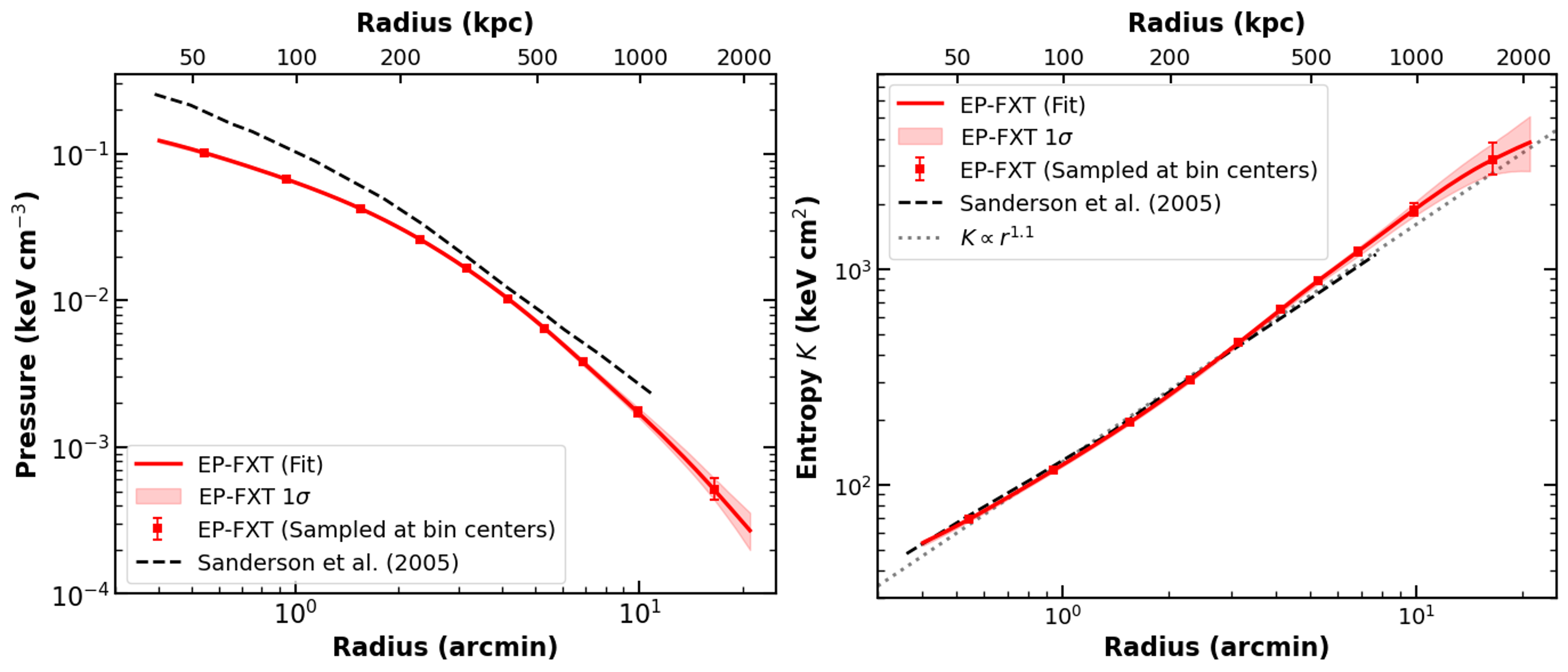}
\caption{Radial distributions of the electron pressure (\textit{left}) and entropy (\textit{right}) of the galaxy cluster A478. The solid red lines show the median profiles derived from the EP-FXT data, the light red shaded regions indicate the $1\sigma$ confidence intervals, and the red data points represent the EP-FXT results sampled at the centers of the radial bins. The dashed black lines show the results of \citet{2005ApJ...630..191S} for comparison. In the right panel, the dotted~gray line indicates the entropy scaling relation expected from the self-similar model dominated by gravitational collapse, $K \propto r^{1.1}$.}
\label{fig:pressure_entropy_profile}
\end{figure*}

   As shown by the electron-pressure profile (left panel of Fig. \ref{fig:pressure_entropy_profile}), the pressure profile of A478 exhibits a continuous, smooth, and monotonic decline from the center outward, reflecting the typical thermal structure of a cool-core galaxy cluster, in which the high-pressure central region gradually transitions to the low-pressure outskirts. A comparison between the pressure profile derived from EP-FXT and that reported by \citet{2005ApJ...630..191S} shows that the two profiles have similar declining radial shapes over the overlapping radial range, indicating that EP-FXT is able to recover the overall radial behavior of the thermal-pressure profile of A478. However, a difference in the absolute normalization remains, with the EP-FXT pressure profile being systematically lower than that of \citet{2005ApJ...630..191S} in the overlapping region. This pressure normalization offset can be primarily understood by considering the combined contributions of temperature and electron density to the pressure measurement. According to Eq.~\ref{eq:pressure}, the electron pressure is proportional to the product of the electron density and temperature. Therefore, differences in the normalization of temperature and density directly affect the derived pressure profile. As shown in Fig.~\ref{fig:specfit}, the temperatures measured by \citet{2005ApJ...630..191S} based on \textit{Chandra} data are generally higher than those obtained from EP-FXT in the inner and intermediate radial regions. Meanwhile, Fig. \ref{fig:ne_fit} shows that the electron-density profile derived from \textit{Chandra} also has a slightly higher normalization than the EP-FXT result. Therefore, the higher temperature and electron-density normalization together naturally lead to a higher pressure normalization in the \citet{2005ApJ...630..191S} profile. In addition, differences in the analysis approaches may also contribute to the pressure normalization offset. \citet{2005ApJ...630..191S} derived the pressure profile using a spectral-mapping deprojection approach based on two-dimensional temperature mapping, whereas in this work the pressure profile is calculated by combining the azimuthally averaged electron-density model with the onion-peeling deprojected temperature profile. Therefore, differences in spatial sampling, deprojection procedures, and pressure reconstruction methods may introduce additional systematic effects. For A478, which has a strong cool core and a steep central surface-brightness gradient, the broader PSF of EP-FXT may smooth the central brightness and temperature structures, thereby affecting the recovery of the central pressure peak. In the outer low-surface-brightness regions, uncertainties in background modeling and temperature deprojection may further affect the pressure estimation. Therefore, the pressure difference between EP-FXT and \citet{2005ApJ...630..191S} is more likely the combined result of temperature and electron-density normalization differences, together with instrumental response, PSF effects, and analysis-method differences, rather than being attributed to a single factor. Overall, the EP-FXT results provide continuous pressure constraints for A478 from the center to nearly the $R_{200}$ region, complementing previous information on the thermal-pressure distribution of this cluster at large radii.

   The entropy profile is shown in the right panel of Fig. \ref{fig:pressure_entropy_profile}. A478 exhibits the characteristic low-entropy core of a cool-core galaxy cluster: the entropy is low in the central region and then increases continuously with radius, reflecting the gradual thermodynamic transition from the dense cool core to the outer main halo. The entropy profile derived from EP-FXT shows a similar radial trend to that reported by \citet{2005ApJ...630..191S} over the overlapping radial range, and is broadly consistent with the self-similar scaling relation expected for gravitational-collapse-dominated evolution, $K \propto r^{1.1}$. Furthermore, EP-FXT extends the entropy measurement to large radii close to $R_{200}$, with the outermost result still increasing with radius, indicating that no obvious flattening of the entropy profile is seen in the outskirts over the radial range probed in this work. Considering the low surface brightness in the outer regions, the outermost entropy value is more sensitive to the background treatment, the deprojected temperature, and the outer-boundary assumption. Therefore, we focus mainly on the overall radial trend of the entropy profile, and do not over-interpret local deviations in any single outer radial bin.

   \subsection{Mass profile}
   \label{sec:obs3.4}
   Under the quasi-static approximation, the hot gas in galaxy clusters can generally be approximated as being in HSE, in which the thermal-pressure gradient of the gas balances the gravitational potential. Based on this assumption, and adopting spherical symmetry, we combined the electron-density model and the deprojected temperature profile derived above to infer the total mass distribution, $M(<r)$, of A478 \citep[e.g.,][]{2013SSRv..177..119E}:
\begin{equation}
M(r) = -\frac{kT(r)r}{G\mu m_{\mathrm{p}}} 
\left( \frac{d\ln n_{\mathrm{e}}(r)}{d\ln r} + \frac{d\ln T(r)}{d\ln r} \right)
\label{eq:mass}
,\end{equation}
where $k$ is the Boltzmann constant, $G$ is the gravitational constant, $\mu=0.609$ is the mean molecular weight of the plasma, $m_{\mathrm p}$ is the proton mass, and $n_{\mathrm e}(r)$ and $T(r)$ are the radial distributions of the electron density and three-dimensional temperature, respectively. The deprojected temperature profile was smoothed using a weighted cubic spline in log $r$--log $T$ space, from which $d\ln T/d\ln r$ was derived for the HSE calculation.

\begin{figure}[htbp]
    \centering
    \includegraphics[width=0.48\textwidth]{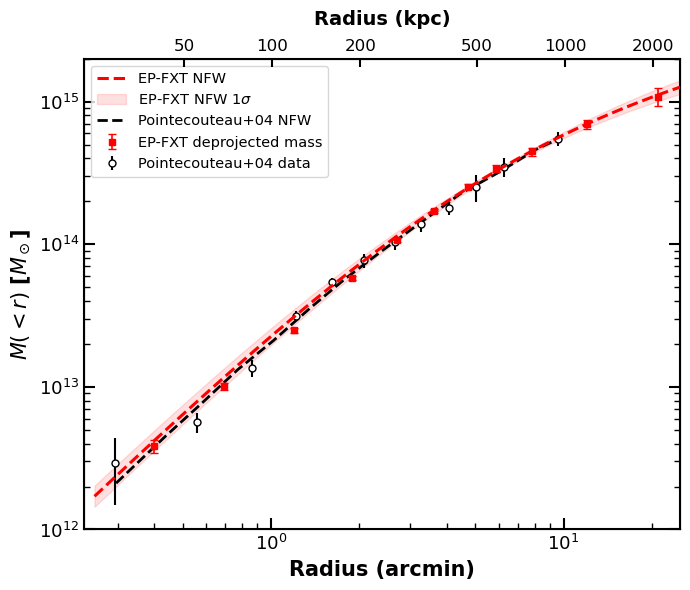}
    \caption{Radial total-mass distribution of A478 and the NFW model fit. The red data points represent the hydrostatic mass derived from the EP-FXT electron-density model and the deprojected temperature profile, the dashed red line the corresponding best-fitting NFW model, and the light red shaded region the $1\sigma$ uncertainty range of the NFW fit. The black open circles represent the mass measurements obtained by \citet{2004A&A...423...33P} based on \textit{XMM-Newton} data, and the dashed black line shows their NFW fit to enable a comparison of the mass distributions over the overlapping radial range.}
    \label{fig:mass_nfw}
\end{figure}

   To propagate the uncertainties in the density and temperature measurements, we used a Monte Carlo approach to construct an ensemble of mass profiles. Specifically, for each set of parameters drawn from the posterior distribution of the density model and each sampled deprojected temperature profile, we calculated the corresponding hydrostatic mass profile. The resulting ensemble was then used to derive the median mass distribution and its $1\sigma$ uncertainty range. Figure \ref{fig:mass_nfw} shows the total mass profile of A478 obtained in this way. The mass increases monotonically with radius, and the overall distribution is smooth, with no obvious unphysical oscillations. Compared with the mass profile derived by \citet{2004A&A...423...33P} based on \textit{XMM-Newton} data, the EP-FXT mass profile is broadly consistent over the overlapping radial range, indicating that the hydrostatic mass estimate obtained from the EP-FXT surface-brightness profile and deprojected temperature profile is reasonable. Furthermore, the EP-FXT result extends the mass constraint to large radii close to $R_{200}$, providing new observational constraints on the mass distribution in the outskirts of A478.
   
   To characterize the dark matter halo structure of A478, we further fitted the mass profile with a Navarro--Frenk--White (NFW) model \citep{1997ApJ...490..493N}. The results show that the mass distribution of A478 can be well described by the NFW profile, yielding $M_{200}=(1.12\pm0.15)\times10^{15}\ \mathrm{M}_{\odot}$ and a concentration parameter of $c=4.58\pm0.86$. The corresponding radius is $R_{200}=2082\pm95\ \mathrm{kpc}$, and the scale radius is $r_{\mathrm{s}}=455\pm109\ \mathrm{kpc}$. The good agreement between the NFW model and the observationally derived mass profile indicates that the large-scale mass distribution of A478 is consistent with the expected dark matter halo structure of massive galaxy clusters in the $\Lambda$ CDM framework. It should be noted that the mass estimate in this work is based on the assumptions of spherical symmetry and HSE, and does not explicitly include nonthermal pressure support from turbulence, cosmic rays, or magnetic fields. Therefore, the resulting mass should be interpreted as the hydrostatic mass estimate derived from the EP-FXT X-ray data.

\section{Thermodynamic properties of the NE and SW sectors}
\label{sec:obs4}
   To further investigate the azimuthal asymmetry of the ICM in A478 and the possible local dynamical processes it may reflect, we selected the NE and SW sectors, where the most prominent thermal-structure differences are seen in the two-dimensional temperature map, for a comparative analysis, with the radial range extending out to 10 arcmin. Compared with the azimuthally averaged radial results presented in the previous section, sector analysis is more sensitive to local deviations in the thermal structure, asymmetries in the gas distribution, and possible dynamical disturbances. Based on the surface-brightness and spectral information in each sector, we derived the radial distributions of the temperature, electron density, thermal pressure, entropy, and hydrostatic mass, and compared their directional differences. This directional analysis is important for understanding the local thermodynamic state of A478. On the one hand, the globally averaged profiles characterize the overall regular structure of the cluster. On the other hand, if systematic differences in the thermal structure are found between different azimuthal directions, they may reflect local imprints left by cold fronts, gas sloshing, or past disturbance events. Therefore, the comparison between the NE and SW sectors not only helps identify inhomogeneous structures within the cluster, but also provides additional constraints for evaluating the validity of the HSE assumption on local scales.

   \subsection{Two-dimensional temperature analysis}
   \label{sec:obs4.1}
\begin{figure}[htbp]
    \centering
    \includegraphics[width=0.4\textwidth]{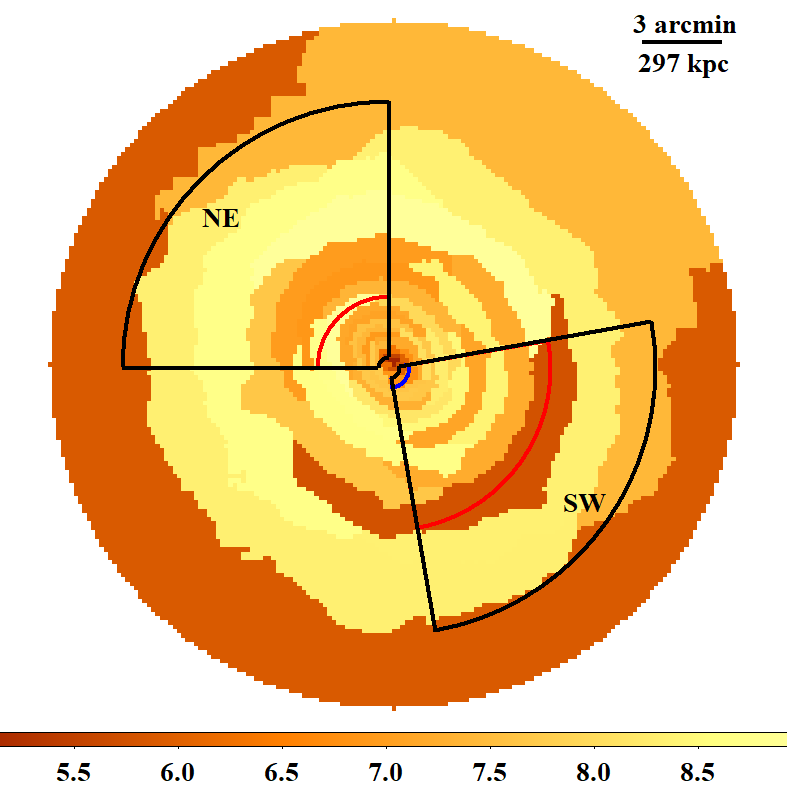}
    \caption{Two-dimensional temperature map of A478 derived from the EP-FXT data using the contour-binning method. The solid black lines mark the NE and SW sectors defined in this work. The red arcs indicate the locations of the temperature discontinuities identified in the subsequent radial analysis. The blue arc in the SW direction marks the position of the inner cold front discovered by \citet{2003ASPC..301...37M} based on \textit{Chandra} observations.}
    \label{fig:temp2d}
\end{figure}

\begin{figure}[htbp]
    \centering
    \includegraphics[width=0.48\textwidth]{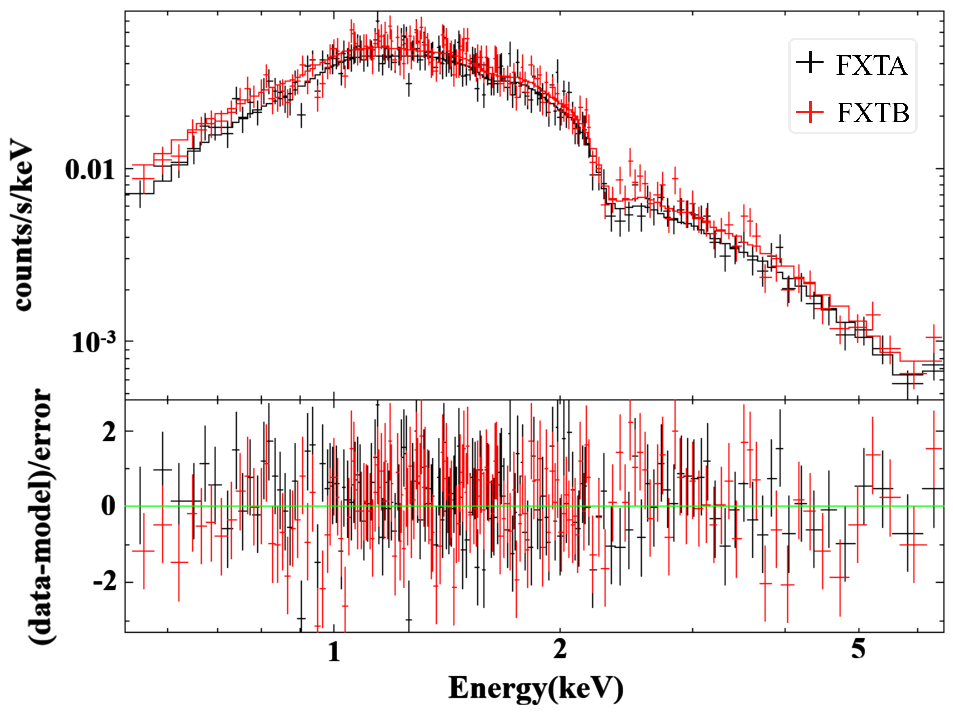}
    \caption{Example of the combined spectrum for a representative \texttt{contour-binning} region in A478. The black and red data points represent the merged FXTA and FXTB spectra, respectively, and the solid lines show the best-fitting model. The lower panel displays the relative residuals. This region is located about 1 arcmin east of the cluster center, and the spectrum has been background-subtracted for both the NXB and CXB components.}
    \label{fig:spectrum_example}
\end{figure}

   To reveal the two-dimensional thermodynamic structure of A478, we applied the \texttt{contour-binning} algorithm proposed by \citet{2006MNRAS.371..829S} to the EP-FXT image for adaptive spatial binning. By requiring each spatial bin to have sufficient statistical quality, we adopted a target signal-to-noise threshold of $S/N = 130$, which resulted in a final division of the image into 44 independent regions. For each region, we extracted the corresponding spectrum and fitted it in the 0.5--7.0 keV band using the \texttt{TBabs} $\times$ \texttt{APEC} model. During the fitting, the redshift was fixed at $z=0.0881$, the hydrogen column density was fixed at $N_{\mathrm H}=2.66\times10^{21}\ \mathrm{cm^{-2}}$ \citep{2004A&A...423...33P}, and the metal abundance was fixed at $0.3Z_{\odot}$. Only the temperature, $kT$, and the model normalization, \texttt{norm}, were allowed to vary freely, in order to obtain a robust spatially resolved temperature structure. The resulting two-dimensional temperature distribution is shown in Fig. \ref{fig:temp2d}. To illustrate the reliability of the spectral fitting for individual regions in the two-dimensional temperature map, we present in Fig. \ref{fig:spectrum_example} an example of the combined spectrum for a representative \texttt{contour-binning} region. This region is located about 1 arcmin east of the cluster center. The figure shows the spectra from the two FXT modules, FXT-A and FXT-B (hereafter FXTA and FXTB), obtained by combining all 14 observations, after subtraction of the NXB and CXB backgrounds. It can be seen that the model provides a good description of the observed data, and the fit residuals do not show any obvious systematic structure, indicating that the two-dimensional temperature measurements presented in this work are reliable.

   As can be seen from the two-dimensional temperature distribution (Fig. \ref{fig:temp2d}), A478 exhibits a relatively uniform cool region in the center ($r \lesssim 2$ arcmin), with temperatures roughly distributed in the range of 5.5--6.5 keV, consistent with the basic properties of a typical cool-core galaxy cluster. Beyond the core, however, the temperature distribution shows a clear azimuthal asymmetry on intermediate scales. In particular, a relatively cool region is present in the SW direction over the radial range $r \approx 4.30$--6.08 arcmin, with a temperature of about 6.4 keV. By contrast, the NE direction at comparable radii exhibits a higher temperature of about 7.5 keV. Based on \textit{XMM-Newton} temperature mapping, \citet{2008A&A...479..307B} showed that the temperature map of A478 is globally regular and approximately elliptical in shape, while still exhibiting a certain degree of non-radial temperature structure on scales of several arcminutes outside the cool core. The EP-FXT temperature map obtained in this work is broadly consistent with this picture, and further reveals a local structure in the SW direction over the range $r \approx 4\arcmin$--$6\arcmin$ that is cooler than the corresponding region in the NE direction. This temperature asymmetry suggests that the ICM in A478 is not fully homogeneous on local scales and may be related to a redistribution of the thermal structure caused by gas sloshing, cold-front structures, or past disturbance events. Motivated by the significant directional differences revealed by this two-dimensional temperature map, in the next subsection we further carry out a quantitative comparison of the radial thermodynamic properties in the NE and SW sectors.

   \subsection{Radial comparison of thermodynamic quantities in the NE and SW sectors}
   \label{sec:obs4.2}
\begin{figure*}[htbp]
\centering
\includegraphics[width=0.85\textwidth]{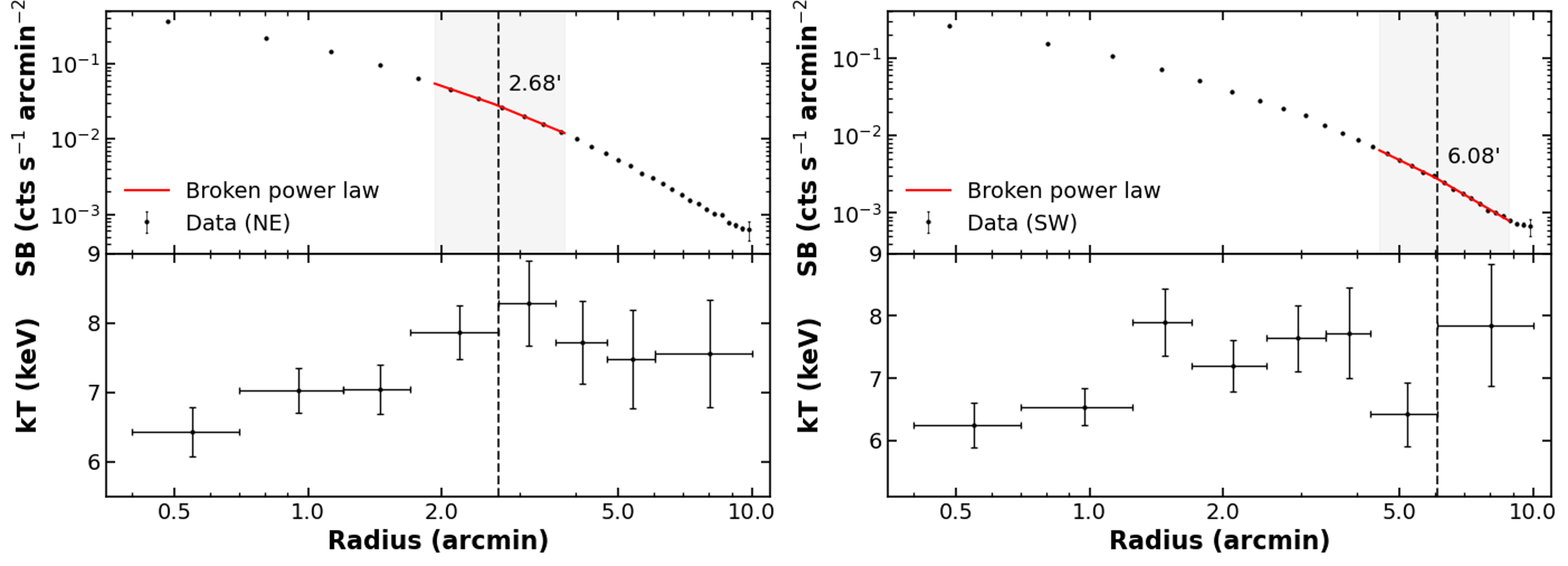}
\caption{Radial distributions of surface brightness (\textit{top}) and temperature (\textit{bottom}) in the NE (\textit{left}) and SW (\textit{right}) sectors of the galaxy cluster A478. In the top panels, the black data points represent the observed surface-brightness profiles, the solid red lines the best-fitting local continuous broken power-law models, the gray shaded regions the radial ranges used for the fitting, and the vertical dashed lines the best-fitting break radii, $R_{\mathrm b}$. For clarity, the single power-law model is not shown; its statistical comparison with the broken power-law model is given in Table \ref{tab:broken_powerlaw}. The bottom panels show the corresponding radial temperature distributions for the two sectors, and the positions of the vertical dashed lines correspond to the red arcs marked in the respective directions in Fig. \ref{fig:temp2d}.}
\label{fig:sector_profiles}
\end{figure*}

   To further investigate the thermodynamic state of the ICM in A478 and its local dynamical properties, we extracted the radial surface-brightness and temperature profiles in the NE ($90^\circ$--$180^\circ$) and SW ($280^\circ$--$370^\circ$) sectors, as indicated by the black sector regions in Fig. \ref{fig:temp2d}. The results are shown in Fig. \ref{fig:sector_profiles}. This analysis extends out to 10 arcmin. The azimuthal angle is defined such that west corresponds to $0^\circ$ and increases counterclockwise. To test for possible local slope changes in the sector surface-brightness profiles, we fitted the same local radial ranges with both a single power-law model and a continuous broken power-law model, and estimated the parameter uncertainties using MCMC sampling. Table~\ref{tab:broken_powerlaw} lists the posterior medians and $1\sigma$ uncertainties of the broken power-law parameters, together with the differences in the Akaike information criterion ($\Delta$AIC) and the Bayesian information criterion ($\Delta$BIC) relative to the single power-law model, which are used to assess whether the introduction of a break provides a statistically improved description of the data.

   Overall, at smaller radii, the surface-brightness and temperature distributions in the NE and SW directions are relatively similar, indicating that the dense gas in the cool-core center of A478 still maintains a relatively regular morphology. At larger radii, however, the two directions begin to show different radial behaviors, revealing local azimuthal thermodynamic asymmetry in A478. In the NE sector, the surface-brightness profile declines smoothly with radius overall, but the local broken power-law fit indicates a slope change at $R_{\mathrm b}=2.68^{+0.28}_{-0.26}$ arcmin (left panel of Fig. \ref{fig:sector_profiles}), with inner and outer power-law slopes of $\alpha_{\mathrm{in}}=2.07^{+0.08}_{-0.16}$ and $\alpha_{\mathrm{out}}=2.44^{+0.12}_{-0.07}$, respectively. Compared with the single power-law model, the broken power-law model significantly improves the fit, with $\Delta\mathrm{AIC}=11.86$ and $\Delta\mathrm{BIC}=12.28$. This radius corresponds to approximately 265 kpc, which is consistent within the uncertainties with the cool-core radius of A478, $r_{\mathrm{cool}}=245\pm66$ kpc, reported by \citet{2024ApJ...968...45N}. Meanwhile, the temperature profile in the NE direction does not show an obvious inverse temperature jump near this radius, but instead increases gradually from the inner to the outer region. Therefore, this surface-brightness break in the NE direction is more likely to correspond to the outer boundary of the cool-core influence region, namely the thermodynamic transition zone between the cooling-dominated gas and the ambient ICM, rather than to a typical cold-front edge.

\begin{table*}[htbp]
\centering
\renewcommand{\arraystretch}{1.3}
\caption{Local broken-power-law fits to the surface-brightness profiles in the NE and SW sectors of A478.}
\label{tab:broken_powerlaw}
\begin{tabular}{lccccccc}
\hline\hline
{Region} & $R_{\mathrm{fit}}$ & $R_{\mathrm b}$ & $I_{\mathrm b}$ & $\alpha_{\mathrm{in}}$ & $\alpha_{\mathrm{out}}$ & $\Delta\mathrm{AIC}$ & $\Delta\mathrm{BIC}$ \\
& (arcmin) & (arcmin) & ($10^{-2}$ cts s$^{-1}$ arcmin$^{-2}$) & & & & \\
\hline
NE & 1.93--3.78 & $2.68^{+0.28}_{-0.26}$ & $2.78^{+0.68}_{-0.56}$ & $2.07^{+0.08}_{-0.16}$ & $2.44^{+0.12}_{-0.07}$ & 11.86 & 12.28 \\
\hline
SW & 4.50--8.82 & $6.08^{+0.31}_{-0.17}$ & $0.28^{+0.03}_{-0.04}$ & $2.83^{+0.10}_{-0.11}$ & $3.36^{+0.10}_{-0.09}$ & 8.94 & 7.81 \\
\hline
\end{tabular}
\tablefoot{
$R_{\mathrm{fit}}$ is the radial range used for the local fit. 
$R_{\mathrm b}$ represents the break radius, $I_{\mathrm b}$ denotes the surface brightness at the break radius, and $\alpha_{\mathrm{in}}$ and $\alpha_{\mathrm{out}}$ are the power-law slopes inside and outside the break radius, respectively. All listed parameters are the medians of the MCMC posterior distributions, and the uncertainties correspond to the $1\sigma$ posterior intervals. The values of $\Delta\mathrm{AIC}$ and $\Delta\mathrm{BIC}$ are defined as $\mathrm{AIC}_{\mathrm{single}}-\mathrm{AIC}_{\mathrm{broken}}$ and $\mathrm{BIC}_{\mathrm{single}}-\mathrm{BIC}_{\mathrm{broken}}$, respectively.
}
\end{table*}

   By contrast, the SW sector exhibits a more complex structural pattern. Its surface-brightness profile shows a clear slope change at $R_{\mathrm b}=6.08^{+0.31}_{-0.17}$ arcmin (right panel of Fig. \ref{fig:sector_profiles}), with inner and outer slopes of $\alpha_{\mathrm{in}}=2.83^{+0.10}_{-0.11}$ and $\alpha_{\mathrm{out}}=3.36^{+0.10}_{-0.09}$, respectively. Compared with the single power-law model, the broken power-law model also provides a better statistical description, with $\Delta\mathrm{AIC}=8.94$ and $\Delta\mathrm{BIC}=7.81$. In the temperature profile, the SW direction contains a local cool region over the radial range $4.30\arcmin$--$6.08\arcmin$, with an average temperature of about 6.4 keV, which is clearly lower than the temperatures in both the adjacent inner and outer regions. This cool region corresponds to the low-temperature structure seen in the two-dimensional temperature map in the SW direction, indicating that it may represent a clump of low-entropy cool gas displaced outward from the core. Combining the surface-brightness and temperature information, the structure near $R_\mathrm b \simeq 6.08\arcmin$ in the SW direction shows several observational features of a cold-front candidate: the surface-brightness profile changes slope near this radius, while the temperature rises from the inner cool region to the hotter ambient gas outside.

   Based on the surface-brightness distributions and projected temperature profiles in the NE and SW sectors, we constructed direction-dependent estimates of the electron density, pressure, entropy, and HSE mass, in order to quantitatively characterize the local thermodynamic asymmetry of A478 and its possible dynamical deviations. The electron density was derived from the sector surface-brightness distribution and described with a double-$\beta$ model. Since the spectral statistics in the individual sectors are not sufficient to support a stable temperature deprojection, we adopted the projected temperature profiles obtained from the spectral fitting in each sector and smoothed them using a weighted cubic spline in log $r$--log $T$ space to derive continuous radial temperature distributions. Therefore, the pressure, entropy, and HSE mass estimates in this section are all calculated using the projected temperatures. They are intended mainly for a relative comparison between the NE and SW directions, and should not be interpreted as strict three-dimensional thermodynamic profiles or independent measurements of the true gravitational mass.

\begin{figure*}[htbp]
\sidecaption
\includegraphics[width=12.5cm]{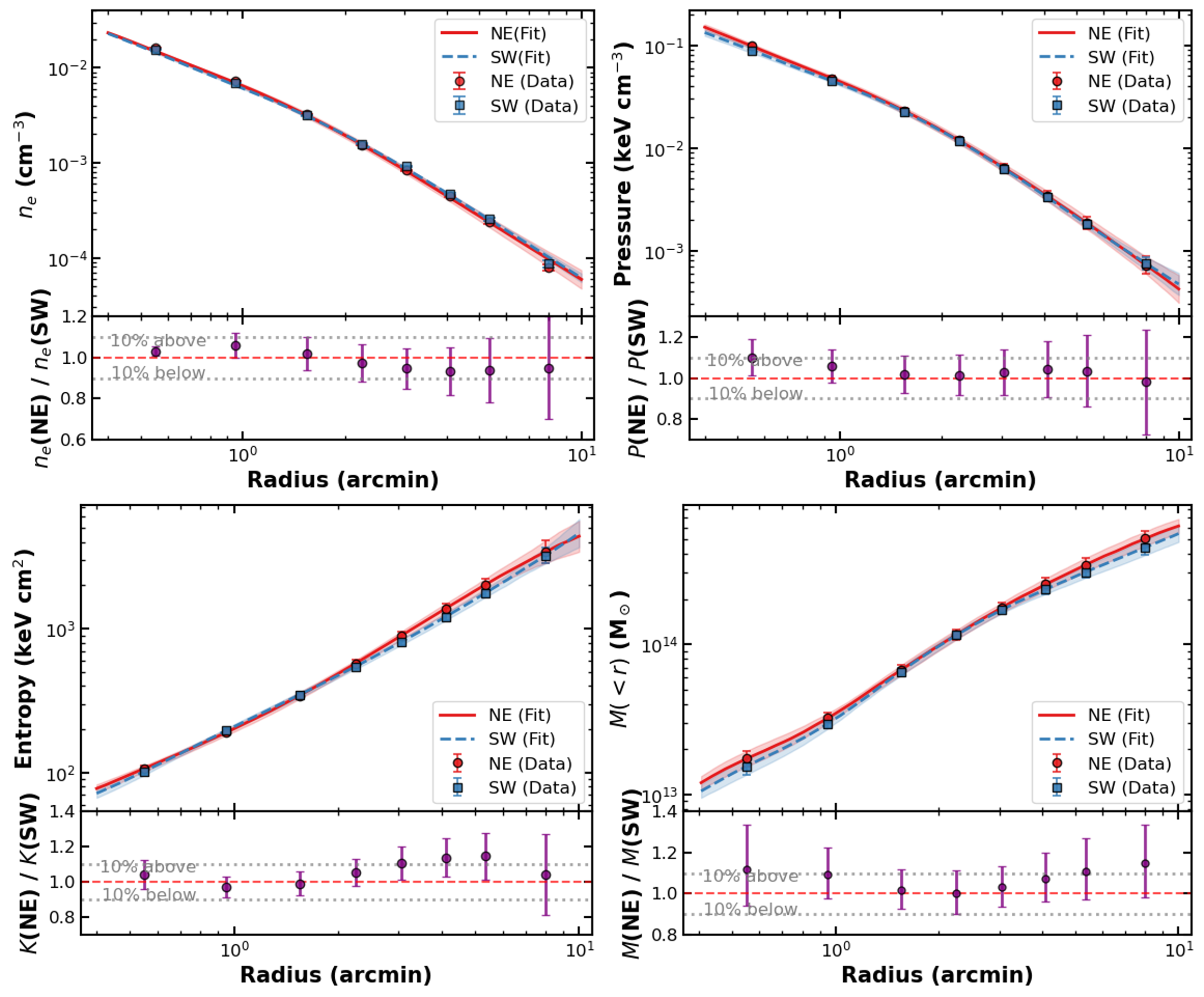}
\caption{Radial distributions of the physical quantities of A478 in the NE (solid red lines) and SW (dashed blue lines) directions. \textit{Top left}: Electron density. \textit{Top right}: Pressure derived from the projected temperature. \textit{Bottom left}: Entropy derived from the projected temperature. \textit{Bottom right}: Mass estimate derived under the HSE assumption. The shaded regions indicate the corresponding $1\sigma$ uncertainty ranges. The lower part of each panel shows the NE/SW ratio, and the two dashed gray lines mark the $\pm10\%$ difference range.}
\label{fig:Four panel diagram}
\end{figure*}

   The results are shown in Fig. \ref{fig:Four panel diagram} for the electron density, pressure, entropy, and HSE mass estimates in the NE and SW direction. Overall, A478 still shows a good degree of similarity between the two sectors over most of the radial range, but the systematic differences seen over several key radial intervals also indicate the impact of local disturbances on the ICM distribution. From the electron-density distribution (top-left panel of Fig. \ref{fig:Four panel diagram}), the NE and SW directions are nearly identical in the inner region, indicating that the gas distribution around the cool-core center remains relatively regular. At larger radii, especially beyond $r \gtrsim 4\arcmin$, the electron density in the NE direction becomes slightly lower than that in the SW direction, with a difference of about 10\%. This pattern of lower density in the NE and higher density in the SW corresponds to the temperature trend described above, in which the NE direction is hotter whereas the SW direction is cooler. By contrast, the pressure distributions (top-right panel of Fig. \ref{fig:Four panel diagram}) show a higher degree of consistency between the two directions. Although the temperature and density in the NE and SW sectors vary in opposite senses, the pressure ratio between the two remains close to unity over the full radial range and stays approximately within $\pm5\%$. No obvious pressure jump is seen even near $r \simeq 6\arcmin$. This indicates that the local structure is more consistent with a contact discontinuity under approximate pressure equilibrium than with a shock structure accompanied by a significant pressure jump. The entropy distribution (bottom-left panel of Fig. \ref{fig:Four panel diagram}) further reveals differences in the thermal history between the two directions. In the intermediate and outer regions at $r \gtrsim 3\arcmin$, the entropy in the NE direction is systematically higher than that in the SW direction, with the ratio reaching about 1.2. This suggests that the SW direction contains relatively low-entropy gas at the same radius, consistent with the low-temperature and high-density features described above. The HSE mass estimates based on the projected temperature (bottom-right panel of Fig. \ref{fig:Four panel diagram}) also show a certain directional difference beyond $r \gtrsim 4\arcmin$, with the mass values in the NE direction being systematically higher than those in the SW direction by about 10\%--15\%. This suggests that the local thermodynamic asymmetry may already affect mass estimates based on the assumption of HSE.

   Taken together, these results show that A478 overall retains the basic properties of a typical relaxed cool-core galaxy cluster, but the SW direction exhibits a set of mutually consistent local thermodynamic features: higher density, lower temperature, approximately continuous pressure, lower entropy, and a lower HSE mass estimate. By contrast, the NE direction is more consistent with a relatively smooth and less disturbed ambient ICM distribution.

\section{Discussion}
\label{sec:obs5}
   A478 has long been regarded as a globally regular strong cool-core galaxy cluster. Its central cooling, smooth surface-brightness distribution, and relatively regular overall thermal structure all support this basic picture. However, the azimuthally resolved analysis presented in this work shows that beneath this globally relaxed appearance, the ICM in A478 is not in an equally uniform thermodynamic state in all directions. In particular, the SW direction exhibits, in the intermediate and outer regions, a combination of structural features including lower temperature, lower entropy, higher density, and nearly continuous pressure, together with observational signatures consistent with an outer cold-front candidate. For systems of this kind, relying only on globally averaged profiles is often insufficient to fully characterize their true dynamical state. Instead, azimuthally resolved thermodynamic information is also required to further clarify the physical origin of local nonequilibrium structures and their potential impact on mass estimates. Motivated by this, in this section we first discuss the local dynamical picture and physical implications reflected by the differences between the NE and SW sectors, and then compare our results with previous studies.

   \subsection{Local thermodynamic asymmetry and its physical interpretation}
   Combining the two-dimensional temperature distribution with the radial thermodynamic results in the NE and SW sectors, we find that although A478 still exhibits the basic properties of a typical cool-core galaxy cluster on global scales, it shows clear thermodynamic asymmetry in specific azimuthal directions. In the central region ($r \lesssim 2\arcmin$), the temperature, density, and pressure distributions are broadly consistent among different directions, indicating that the cool-core center still maintains a relatively regular structure overall. In the intermediate and outer regions, however, the NE and SW directions show systematic differences, suggesting that local dynamical processes have affected the ICM distribution to some extent. In particular, the SW direction exhibits a relatively low-temperature, low-entropy, and high-density structure over the range $r \approx 4\arcmin$--$6\arcmin$, and shows a density drop, a temperature rise, and an approximately continuous pressure profile near $R_{\mathrm b}=6.08^{+0.31}_{-0.17}\ \mathrm{arcmin}$. Taken together, these observational features are broadly consistent with the picture of a cold-front candidate, indicating that the SW direction of A478 may host a locally disturbed region driven by gas sloshing.

   Hydrodynamical simulations have shown that when a secondary subhalo passes the main cluster at relatively large radii, it can perturb the gravitational potential well of the primary system, causing the low-entropy, high-density cool gas originally located in the central deep potential to undergo bulk sloshing and gradually extend outward into a curled, spiral-like cool-gas structure \citep{2006ApJ...650..102A, 2007PhR...443....1M}. Because the sloshing cool gas generally has a complex three-dimensional geometry, and because its projected appearance depends on the line of sight, the presence of multiple cold-front interfaces within the same system, even at different radii along similar azimuthal directions, is therefore expected \citep{2011MNRAS.413.2057R}. Similar multiple cold fronts and their associated large-scale sloshing structures have been reported in several cool-core clusters. For example, the Perseus cluster hosts large-scale sloshing cold-front interfaces extending from the core to the outer regions \citep{2012ApJ...757..182S,2018NatAs...2..292W}, while observational and simulation studies of Abell~496 have shown that gas sloshing triggered by a minor merger can produce multiple cold fronts at different radii within the same system \citep{2012MNRAS.420.3632R,2014A&A...570A.117G}. In the case of A478, \textit{Chandra} observations have already identified an inner cold front at about 60 kpc, or approximately 0.6 arcmin, in the SW direction \citep{2003ASPC..301...37M}. In this work, we identify a new outer cold-front candidate at a larger radius along a broadly similar azimuthal direction, at $R_{\mathrm b}=6.08^{+0.31}_{-0.17}\ \mathrm{arcmin}$. Combined with the low-temperature, low-entropy, high-density, and approximately pressure-continuous properties of this region, the local thermal structure in the SW direction of A478 is likely related to gas sloshing. This suggests that the system experienced a relatively mild disturbance event that did not significantly disrupt its overall cool-core structure, but nevertheless left observable local thermodynamic imprints in the ICM.

   By contrast, the NE direction appears much smoother in both the two-dimensional temperature map and the radial distributions of the derived physical quantities, with its temperature, entropy, and HSE mass profiles being more consistent with the expectation for a relaxed state in the global average sense. The comparison between the NE and SW directions further indicates that although approximate local pressure equilibrium is still maintained near the cold-front candidate, local dynamical disturbances may already have affected the mass estimates based on the assumption of HSE, causing the SW direction to show lower HSE mass estimates at intermediate and outer radii \citep{2007ApJ...655...98N, 2009ApJ...705.1129L}. A478 is therefore better understood as a cool-core cluster that is globally relaxed but locally disturbed: its central cool-core structure remains relatively stable, while the ICM in some azimuthal directions at larger radii still preserves thermodynamic imprints left by gas sloshing.
   
   This result indicates that even in cool-core galaxy clusters with globally regular morphologies, local asymmetric structures may still have a non-negligible impact on thermodynamic analyses and mass estimates. Therefore, for such systems, azimuthally resolved thermodynamic studies are a necessary complement to globally averaged radial analyses for understanding their local dynamical state and assessing the reliability of HSE-based mass estimates.

   \subsection{Comparison with previous studies}
   Previous studies have already provided a relatively comprehensive picture of A478 as a classical strong cool-core galaxy cluster. In particular, \citet{2004A&A...423...33P} performed a systematic analysis of the gas and total mass distributions of A478 based on \textit{XMM-Newton} observations, extending the radial study to about $0.5R_{200}$ and characterizing the thermodynamic and mass distributions of the cluster from the center to intermediate radii. Our results are consistent with this basic picture: A478 overall exhibits a relatively regular cool-core structure, and its mass distribution can be well described by an NFW model. Building on this foundation, we use homogeneous EP-FXT data, together with a uniform instrumental response and analysis procedure, to extend the measurements of the temperature, electron density, pressure, entropy, and HSE mass estimates to nearly the $R_{200}$ scale. This allows us to continuously characterize the thermodynamic properties of A478 from the cool-core center to the outskirts within a single observational framework.

   On the other hand, the two-dimensional temperature study of \citet{2008A&A...479..307B} had already suggested that although A478 is globally regular in morphology, the regions outside the cool core are not fully homogeneous in different azimuthal directions, but instead show local temperature inhomogeneities. Our results are consistent with this overall picture, and further use the EP-FXT data to perform a quantitative azimuthally resolved thermodynamic analysis of the NE and SW directions. Through a radial comparison between the NE and SW sectors, we show that the SW direction contains a relatively low-temperature, low-entropy, and high-density structure over the range $r \approx 4\arcmin$--$6\arcmin$, and identify an outer cold-front candidate along the same direction. Near its outer boundary, this structure shows a density drop, a temperature rise, and an approximately continuous pressure profile, consistent with the thermodynamic characteristics expected for a cold-front candidate. In this sense, the present work extends the local thermal-structure differences suggested by previous two-dimensional temperature maps into quantitatively resolved azimuthal thermodynamic profiles, and provides more specific observational constraints for understanding the local ICM inhomogeneity of A478 and its possible dynamical origin.

   Compared with the studies of \citet{2014PASJ...66...99O,2016MNRAS.456.4475O}, the focus of the present work is also different. \citet{2014PASJ...66...99O} included A478 in a joint \textit{Suzaku}+Subaru sample of four relaxed clusters, mainly discussing the average thermodynamic properties of the outer ICM and the radial variation of the hydrostatic-to-lensing mass ratio. \citet{2016MNRAS.456.4475O} further constrained the central mass distribution of A478 from the perspectives of weak lensing and the stellar luminosity distribution. These studies provide important background for understanding the mass structure of A478 and the HSE bias in the outskirts, and also offer valuable reference points for interpreting our results. It is worth noting that the weak-lensing analysis of \citet{2016MNRAS.456.4475O}, based on a single NFW model, yielded $M_{200}=12.93^{+3.96}_{-3.22}\times10^{14}\ h_{70}^{-1}\ \mathrm{M}_{\odot}$. This central value is slightly higher than the $M_{200}=(1.12\pm0.15)\times10^{15}\ \mathrm{M}_{\odot}$ obtained in this work from the EP-FXT data under the HSE assumption, but the two measurements remain consistent within the uncertainties. Given that weak-lensing and HSE mass estimates rely on different methods and are subject to different sources of systematic uncertainty, such a difference is not unexpected. Instead, it highlights the complementary constraints provided by the two approaches on the mass structure of A478.

   Overall, the present work shows good continuity and complementarity with previous studies. Based on homogeneous EP-FXT data and using a uniform instrumental response and analysis procedure, we continuously characterize the thermodynamic properties of A478 from the cool-core center to nearly the $R_{200}$ scale, thereby providing more complete observational constraints on the ICM structure over a large radial range. At the same time, we further quantify the two-dimensional temperature inhomogeneity suggested by previous studies into an azimuthally resolved thermodynamic asymmetry. In particular, we identify a larger-scale low-temperature and low-entropy structure in the SW direction, together with an associated outer cold-front candidate. These results indicate that even in a globally regular cool-core galaxy cluster, local dynamical disturbances can leave persistent nonequilibrium thermodynamic imprints in specific azimuthal directions and may have potential systematic effects on mass estimates based on the HSE assumption.

\section{Conclusions}
\label{sec:obs6}
Using multiple on-axis observations of the galaxy cluster A478 performed with EP-FXT in 2024, we carried out a systematic analysis of its X-ray morphology, radial thermodynamic distributions, and azimuthal asymmetries. Benefiting from the combined advantages of the large field of view, low background, and strong imaging spectroscopic capability of EP-FXT, we extended the measurements of the temperature, electron density, pressure, entropy, and total mass of A478 out to$R_{200}$. This enabled us to characterize, in a relatively complete manner, the thermal structure of the cluster from its cool-core center to the outskirts, and to provide new observational constraints on its local dynamical disturbances.

Based on the globally averaged results, A478 overall appears to be a typical strong cool-core galaxy cluster. Its X-ray surface-brightness distribution is regular, strongly peaked in the center, and described  well by a double-$\beta$ model. The annular spectral analysis shows that its temperature profile has the characteristic behavior of a cool-core cluster, with a low central temperature, a radial increase toward intermediate radii, and a gradual decline in the outskirts. Over the radial range overlapping with previous \textit{XMM-Newton} measurements, the EP-FXT temperature results are broadly consistent with the literature values and further extend the temperature measurement to $R_{200}$. On this basis, the electron-density profile derived from the deprojected surface-brightness distribution, together with the pressure and entropy profiles obtained by combining this density profile with the deprojected temperature distribution, all show continuous structures consistent with a relaxed cool-core system. Under the assumption of HSE, the derived total mass distribution can be well fitted by an NFW model, yielding $M_{200}=(1.12\pm0.15)\times10^{15}\ \mathrm{M}_{\odot}$, $c=4.58\pm0.86$, and $R_{200}=2082\pm95\ \mathrm{kpc}$. These results indicate that the large-scale mass structure of A478 is consistent with the expected dark matter halo structure in the $\Lambda$CDM framework.

Although A478 exhibits a relatively regular strong cool-core structure on global scales, there are local thermodynamic directional differences in the NE and SW sectors over the radial range analyzed in this work. Both the two-dimensional temperature map and the sector radial profiles show that the SW direction contains a relatively low-temperature, low-entropy, and high-density gas structure over the range $r \sim 4\arcmin$--$6\arcmin$. Near its outer boundary, at $R_{\mathrm b}=6.08^{+0.31}_{-0.17}\ \mathrm{arcmin}$, this structure shows a density drop, a temperature rise, and an approximately continuous pressure profile. This combination of properties is broadly consistent with an outer cold-front candidate and is consistent with the presence of a locally disturbed region driven by gas sloshing in this direction. By contrast, the temperature, entropy, and HSE mass estimates in the NE direction are smoother and closer to the expectation for a relaxed state in the global average sense. Therefore, A478 is better understood as a cool-core cluster that is globally relaxed but locally disturbed.

The azimuthally resolved analysis also shows that local thermodynamic differences can affect mass estimates based on the assumption of HSE. We find that the HSE mass estimates in the SW sector, derived using the projected temperature, are systematically lower than those in the NE direction at intermediate and outer radii, with a difference of about 10\%--15\%. Considering that the sector mass profiles are used mainly for relative comparisons between azimuthal directions, rather than as independent measurements of the true gravitational mass, this difference more likely reflects the systematic impact of local dynamical disturbances on HSE-based mass estimates. In other words, even in galaxy clusters with globally regular morphologies that are usually classified as strong cool-core systems, local nonequilibrium processes may still introduce non-negligible systematic biases into thermodynamic analyses and mass estimates.

Overall, A478 maintains a globally regular cool-core structure and a quasi-static thermodynamic distribution dominated by its large-scale gravitational potential while still showing local azimuthal thermal asymmetries that may be driven by gas sloshing. This work not only extends the measurements of multiple thermodynamic quantities in A478 to $R_{200}$, but also demonstrates the capability of EP-FXT to study the low-surface-brightness outskirts of galaxy clusters and their subtle dynamical structures. With future deeper observations and sample-based analyses, EP-FXT is expected to play an important role in revealing the thermodynamic evolution of cool-core clusters from their cores to their outskirts, and in assessing the impact of local dynamical processes on mass estimates.

\begin{acknowledgements}
\textit{Einstein} Probe (EP) is a space mission supported by the Strategic Priority Program on Space Science of the Chinese Academy of Sciences, in collaboration with ESA, MPE, and CNES (Grant Nos. XDA15310000, XDA15310103, and XDA15052100). This work was supported by the EP Program, which is funded by the Strategic Priority Research Program of the Chinese Academy of Sciences (Grant No. XDA15310303), the National Key R\&D Program of China (Grant No. 2025YFF0511104), the International Partnership Program of the Chinese Academy of Sciences (Grant No. 013GJHZ2024015FN), and the Postgraduate Scientific Research Innovation Project of Xiangtan University (Grant No. XDCX2025Y267). H.Y. acknowledges support from the National Natural Science Foundation of China (Grant No. 12573003). A.L. acknowledges support from the National Natural Science Foundation of China (Grant No. 12588202) and the China Manned Space Program (Grant No. CMS-CSST-2025-A04).
\end{acknowledgements}

\bibliographystyle{aa} 
\bibliography{references}

\clearpage
\onecolumn

\begin{appendix}
\section{EP-FXT observations of A478}
\label{app:observations}

Table~\ref{tab:obslog} lists the individual EP-FXT observations of A478
used in this work. Details of the observations and data reduction are given in Sect.~\ref{sec:observations}.

\begin{table}[htbp]
\caption{EP-FXT observations of A478.}
\label{tab:obslog}
\centering
\small
\setlength{\tabcolsep}{5pt}
\renewcommand{\arraystretch}{0.95}
\begin{tabular}{ccccccc}
\hline\hline
ObsID & Date & RA (deg) & Dec (deg) & Exposure (ks) & Filter & Instrument \\
\hline
11904195603 & 2024-10-26 & 63.37 & 10.49 & 10.29 & thin   & A/B \\
11904195604 & 2024-10-26 & 63.37 & 10.49 & 8.78  & thin   & A/B \\
11904195605 & 2024-10-27 & 63.37 & 10.49 & 11.51 & thin   & A/B \\
11904195677 & 2024-10-30 & 63.37 & 10.49 & 21.43 & thin   & A/B \\
11904195678 & 2024-10-30 & 63.37 & 10.49 & 6.10  & thin   & A/B \\
11904195679 & 2024-10-31 & 63.38 & 10.49 & 15.06 & medium & A/B \\
11904195680 & 2024-10-31 & 63.38 & 10.49 & 18.04 & medium & A/B \\
11904195681 & 2024-11-02 & 63.38 & 10.49 & 9.64  & medium & A/B \\
11904195682 & 2024-11-03 & 63.38 & 10.49 & 20.17 & medium & A/B \\
11904195683 & 2024-11-03 & 63.38 & 10.49 & 18.21 & medium & A/B \\
11904195780 & 2024-12-03 & 63.37 & 10.43 & 11.92 & thin   & A/B \\
11904195781 & 2024-12-04 & 63.37 & 10.43 & 17.64 & thin   & A/B \\
11904195782 & 2024-12-05 & 63.37 & 10.43 & 10.57 & thin   & A/B \\
11904195783 & 2024-12-06 & 63.37 & 10.43 & 17.70 & thin   & A/B \\
\hline
\end{tabular}
\end{table}
\FloatBarrier

\section{Posterior distributions of the double-$\beta$ model parameters for the electron-density profile}
\label{app:density_corner}
To illustrate the uncertainties and correlations of the fitted parameters in the double-$\beta$ model for the electron-density profile, we constructed a corner plot based on the MCMC sampling results. The plot includes the one-dimensional marginalized posterior distributions of the free parameters and the two-dimensional joint posterior distributions for all parameter pairs, allowing a direct assessment of the strength of the parameter constraints, the directions of degeneracy, and the overall posterior morphology. The free parameters adopted in this work are $\beta_1$, $r_{\mathrm{c}1}$, and $n_{01}$ for the inner component, and $\beta_2$, $r_{\mathrm{c}2}$, and $n_{02}$ for the outer component.

\begin{figure}[htbp]
    \centering
    \includegraphics[width=0.85\textwidth]{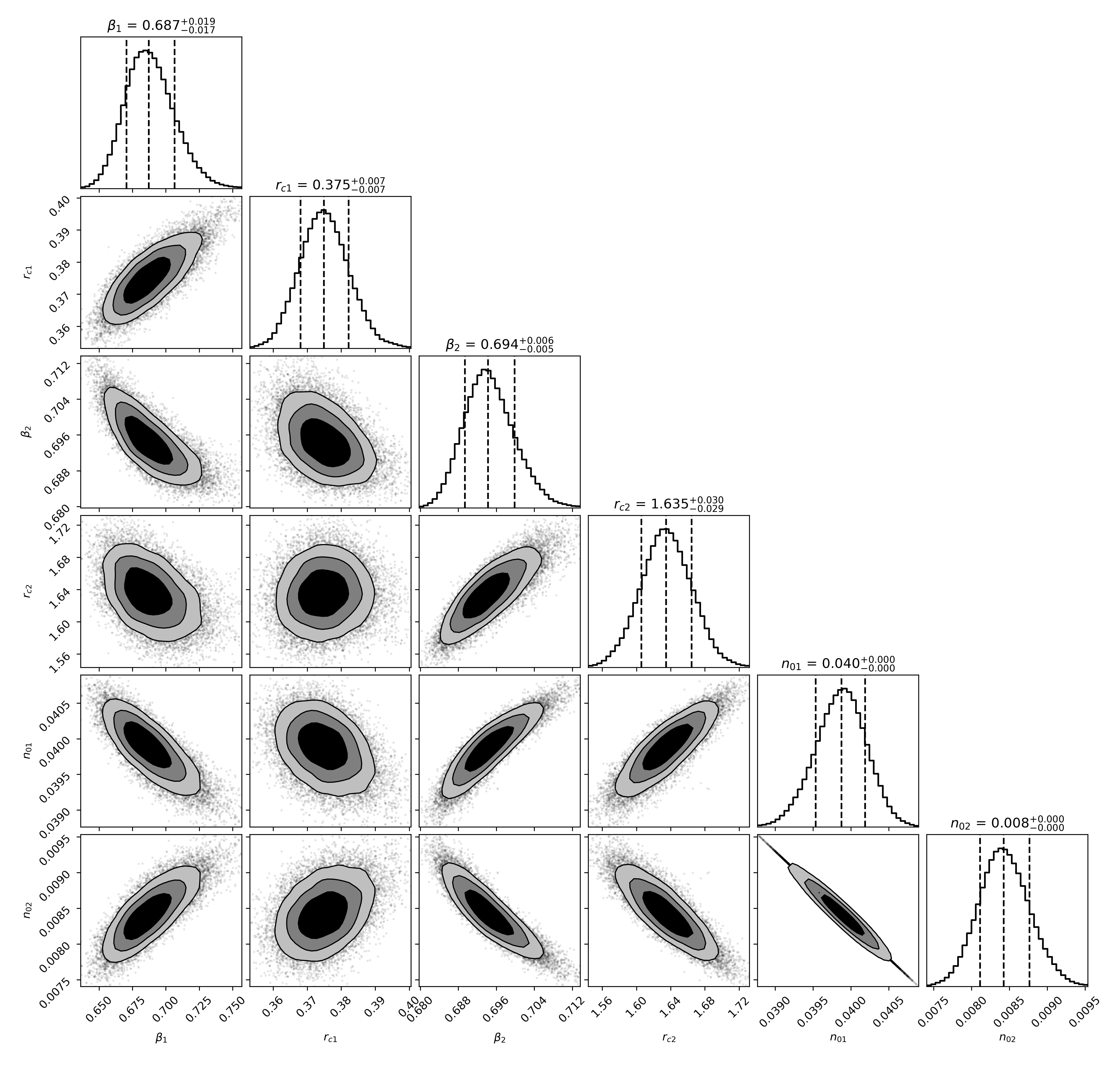}
    \caption{Corner plot of the posterior distributions of the double-$\beta$ model parameters for the electron-density profile of A478. The diagonal panels show the one-dimensional marginalized posterior distributions of $\beta_1$, $r_{\mathrm{c}1}$, $\beta_2$, $r_{\mathrm{c}2}$, $n_{01}$, and $n_{02}$. The parameter values given in the panel titles correspond to the 50th percentiles of the posterior distributions, while the upper and lower uncertainties are defined by the deviations of the 84th and 16th percentiles from the median, respectively. The off-diagonal panels display the two-dimensional joint posterior distributions of the parameter pairs and their correlations. The shaded contours correspond, from inner to outer, to cumulative posterior probabilities of 39.3\%, 67.5\%, and 86.4\%; for a two-dimensional Gaussian distribution, these approximately correspond to the 1$\sigma$, 1.5$\sigma$, and 2$\sigma$ credible regions, respectively. This figure is intended to illustrate the statistical uncertainties of the double-$\beta$ model parameters and the couplings among them.}
    \label{fig:density_corner}
\end{figure}

In Fig. \ref{fig:density_corner}, the titles of the diagonal panels give the 50th percentiles of the parameters together with the corresponding 16th and 84th percentile intervals, which are the source of the best-fitting values and $1\sigma$ uncertainties listed in Table \ref{tab:ne_fit}. The shaded contours in the off-diagonal panels represent the cumulative credible regions of the two-dimensional joint posterior distributions; according to the plotting setup, the three contour levels from inner to outer correspond to posterior probabilities of 39.3\%, 67.5\%, and 86.4\%, respectively. To improve the readability of the figure, both the one-dimensional and two-dimensional posterior distributions were smoothed moderately. The corner plot shows that some of the parameters in the double-$\beta$ model are clearly correlated, in particular the typical coupling between the central electron density, the core radius, and the slope parameter. Such correlations reflect the intrinsic degeneracies of a multiparameter model when fitting the electron-density profile. Overall, however, the posterior distributions of all parameters are relatively concentrated and display well-behaved unimodal structures, indicating that the parameter constraints of the electron-density model obtained in this work are stable.

\FloatBarrier

\end{appendix}

\end{document}